\documentstyle[preprint,revtex]{aps}
\begin{document}
\draft
\begin{title}
Quantum Gauge Equivalence in QED
\end{title}
\author{K.~Haller and E.~Lim-Lombridas}
\begin{instit}
Department of Physics, University of Connecticut, Storrs, Connecticut
06269
\end{instit}
\begin{abstract}
We discuss gauge transformations in QED coupled to a charged spinor
field, and examine whether we can gauge-transform the entire formulation
of the theory from one gauge to another, so that not only the gauge and
spinor fields, but also the forms of the operator-valued Hamiltonians
are transformed. The discussion includes the covariant gauge, in which
the gauge condition and Gauss's law are not primary constraints on
operator-valued quantities; it also includes the Coulomb gauge, and the
spatial axial gauge, in which the constraints are imposed on operator-
valued fields by applying the Dirac-Bergmann procedure. We show how to
transform the covariant, Coulomb and spatial axial gauges to what we
call ``common form,'' in which all particle excitation modes have
identical properties. We also show that, once that common form has been
reached, QED in different gauges has a common time-evolution operator
that defines time-translation for states that represent systems of
electrons and photons. By combining gauge transformations with changes
of representation from standard to common form, the entire apparatus of
a gauge theory can be transformed from one gauge to another.
\end{abstract}

\section{Introduction} Quantum Electrodynamics (QED) embodies the
problems characteristic of a field theory with an infinite number of
degrees of freedom, as well as those that attach to a theory with time-
independent constraints. Thus, QED manifests the divergent, albeit
renormalizable radiative corrections which also characterize other field
theories, many of which do not include any gauge fields and are not
invariant to any local gauge transformations. QED is also a gauge
theory, and must obey a time-independent constraint which imposes
relations among fields not only while different particles are in
interaction with each other, but also in the so-called ``asymptotic
limit,'' when all particles are far apart and not interacting. Such
constraints can conflict with idealized models in which, when two
different varieties of particles recede from each other, field
interaction effectively cease, and their corresponding asymptotic ``in''
or ``out'' fields commute. We note, for example, that the ``LSZ''
axioms, which were believed to be general enough to accommodate all
field theories, \cite{lehmann} are not fully compatible with QED because
canonical commutation rules between asymptotic ``in'' or ``out'' Dirac
and Maxwell fields are incompatible with Gauss's law. \cite{khallerd18}

Because it is difficult to graft the constraints that characterize QED
onto the apparatus that generates perturbative $S$-matrix elements,
gauge-invariance is not nearly as straightforward in QED as it is in
classical electrodynamics. On a superficial level, gauge-invariance in
QED is equated to the very well established ``test of gauge-
invariance,'' which requires that $S$-matrix elements vanish when they
have external photon lines polarized in the $k_\mu$ direction.
\cite{bjorken} This principle is beyond dispute when applied to tree
graphs. It also applies to graphs with radiative corrections, although
apparent violations can be caused by regularization procedures that are
incompatible with it. Furthermore, we can raise the question of whether
$S$-matrix elements are identical in different gauges. Here, again, this
identity is well established for tree graphs. In graphs with radiative
corrections, the renormalization procedure has not been firmly
established since the finite parts of divergent are integrals not well
defined in non-covariant gauges, such as axial gauges. \cite{leibbrandt}

In classical electrodynamics, gauge transformations can be implemented
simply by adding the gradient $\partial_\mu\chi$ to $A_\mu$. In QED,
various approaches may be taken. It is possible to consider $\chi$ to be
a $c$-number. In that case, the gauge transformation cannot readily
shift the formulation of QED from one gauge to another, since different
representations of the operator-valued fields and Hamiltonians will
generally be required in different gauges. We can also consider $\chi$
to be an operator-valued field. In that case, $\partial_\mu\chi$ is not
well defined, since the time derivative operator is
$\partial_0\chi=i[H,\chi]$, and the gauge-dependent Hamiltonian presents
an obstacle to the unambiguous definition of time evolution. It would be
desirable to be able to demonstrate identical time displacement, in
different gauges, of a state vector that represents an observable
state---for example, a system of electrons and photons. Such a
development might not make any substantial difference in the task of
calculating $S$-matrix elements or energy levels of quasi-bound states.
But it might contribute to our fundamental understanding of the theory.

In this paper, we will address this problem, and review the theoretical
apparatus required for its consideration.

\section{QED in covariant gauges} \label{sec:QEDincovrianggauges} The
most common and familiar example of a gauge theory is Quantum
Electrodynamics (QED), in a manifestly covariant gauge.  In a manifestly
covariant gauge (or ``covariant gauge,'' for short), Lorentz
transformations do not change the gauge; in other gauges, after a
Lorentz boost, a subsequent gauge transformation is required to return
to the original gauge.  A useful Lagrangian for QED in a manifestly
covariant gauge is \cite{lautrup} \begin{equation} {\cal L} = -\case 1/4
F_{\mu\nu}F^{\mu\nu} - j_\mu A^\mu - G(\partial_\mu A^\mu) + \case 1/2
(1-\gamma)G^2 + \bar{\psi}(i\gamma_\mu\partial^\mu - m)\psi
\end{equation} where $F_{\mu\nu} = \partial_\mu A_\nu - \partial_\nu
A_\mu$ and $j^\mu = e\bar{\psi}\gamma^\mu\psi$. Here $G$ is the so-
called gauge-fixing field, and $\gamma$ is a parameter for ``tuning''
the covariant gauge to various possible alternatives---the Feynman
\cite{feynman}, Landau \cite{landau}, Fried-Yennie \cite{fried}, or
other variants.  The Euler-Lagrange equations for the gauge fields
derived from this Lagrangian are \begin{equation} \partial_\nu
F^{\mu\nu} + j^\mu - \partial^\mu G = 0; \end{equation} it is
instructive to rewrite these equations as \begin{equation}
\partial_0{\bf E} - \nabla \times {\bf B} + {\bf j} = -\nabla G
\label{eq:maxwellampere} \end{equation} and \begin{equation} \nabla
\cdot {\bf E} - j_0 = -\partial_0 G. \label{eq:gauss} \label{eq:DoG}
\end{equation} Further Euler-Lagrange equations are \begin{equation}
\partial_\mu A^\mu = (1-\gamma)G, \end{equation} \begin{equation}
\partial_\mu\partial^\mu G = 0, \end{equation} and \begin{equation}
(i\gamma_\mu D^\mu - m)\psi = 0 \end{equation} where $D_\mu =
\partial_\mu + ieA_\mu$.  The Lagrangian, ${\cal L}$, also determines
the momenta canonical to the gauge fields, which are given by
\begin{equation} \Pi_i = \frac{\delta {\cal L}}{\delta (\partial^0A_i)}
= -F_{0i} \label{eq:Pii} \end{equation} and \begin{equation} \Pi_0 =
\frac{\delta {\cal L}}{\delta (\partial^0A^0)} = -G; \label{eq:Pi0}
\end{equation} where $\Pi_i$ is a component of ${\bf \Pi} = -{\bf E}$,
and ${\bf E}$ is the electric field.

The momentum conjugate to $\psi$ is $i\psi^\dagger$.  We observe that
the introduction of the gauge-fixing field $G$ has provided us with a
momentum conjugate to $A^0$; this avoids the necessity of imposing
primary constraints on operator-valued fields, and allows us to preserve
the canonical commutation (or anticommutation) relations among all the
participating field operators.  On the other hand, as shown in
Eqs.~(\ref{eq:maxwellampere}) and (\ref{eq:gauss}), the gauge-fixing
field also adds spurious terms to Maxwell's equations.  In order to
guarantee that this theory is really QED, we will ultimately have to
prevent the derivatives of $G$ from affecting the validity of Maxwell's
equations. First, however, we proceed with the development of the
canonical apparatus, and impose the equal-time commutation (and
anticommutation) rules.  Because there are no primary operator-valued
constraints in this formulation, the following completely canonical
rules apply \begin{equation} [A_i({\bf x}), \Pi_j({\bf y})] =
i\delta_{ij}\delta({\bf x-y}), \label{eq:comAiPij} \end{equation}
\begin{equation} [A_0({\bf x}),G({\bf y})] = -i\delta({\bf x-y})
\label{eq:comAoG} \end{equation} and \begin{equation}
\{\psi^\dagger({\bf x}),\psi({\bf y})\} = \delta({\bf x-y}).
\label{eq:anticompsi} \end{equation} We evaluate the Hamiltonian for the
covariant gauge, $H^{\mbox{\rm\scriptsize cov}} = \int d{\bf x}\ {\cal
H}^{\mbox{\rm\scriptsize cov}}({\bf x})$, with \begin{equation} {\cal
H}^{\mbox{\rm\scriptsize cov}} = \partial^0A^i\frac{\delta {\cal
L}}{\delta(\partial^0 A^i)} + \partial^0A^0\frac{\delta {\cal
L}}{\delta(\partial^0 A^0)} + \partial^0\psi\frac{\delta {\cal
L}}{\delta(\partial^0 \psi)} - {\cal L}; \label{eq:hamilden}
\end{equation} and we represent ${\cal H}^{\mbox{\rm\scriptsize cov}}$
as ${\cal H}^{\mbox{\rm\scriptsize cov}} = {\cal
H}_0{}^{\mbox{\rm\scriptsize cov}} + {\cal H}_{\mbox{\rm\scriptsize
I}}{}^{\mbox{\rm\scriptsize cov}}$, where\footnote{Equation
(\ref{eq:Hcal0cov}) assumes implicit integration by parts.}
\begin{eqnarray} {\cal H}_0{}^{\mbox{\rm\scriptsize cov}} &=& \case 1/2
{\bf \Pi \cdot \Pi} + \case 1/4 F_{ij}F^{ij} + G\nabla\cdot{\bf A} +
A^0\nabla\cdot{\bf \Pi} - \case 1/2(1-\gamma)G^2\nonumber\\
&&\mbox{\hspace{1in}}+\ \psi^\dagger(-i\mbox{\boldmath
$\alpha$}\cdot\nabla + \beta m)\psi \label{eq:Hcal0cov} \end{eqnarray}
and \begin{equation} {\cal H}_{\mbox{\rm\scriptsize
I}}{}^{\mbox{\rm\scriptsize cov}} = j_0A^0 - {\bf j \cdot A};
\label{eq:HIdensity} \end{equation} we use $\mbox{\boldmath
$\alpha$}=\gamma^0{\mbox{\boldmath $\gamma$}}$ and $\beta = \gamma^0$.

In order to obtain a Fock space of particle states, we represent the
gauge field in terms of creation and annihilation operators for photons
and electrons.  There are, however, more independent degrees of freedom
in the gauge fields than we can accommodate with the two helicity modes
available for propagating photons; additional operator-valued excitation
modes are necessary to represent the quantized gauge field.  These
additional modes must neither carry energy-momentum, nor have any
probability of being observed, so that the theory's unitarity is
preserved within the space determined by the electrons and the two
helicity modes of observable photons.  In order to accommodate these
requirements consistently, we make use of ``ghost modes,'' i.e.,
excitation modes of the gauge fields which create zero-norm states.  We
will designate the annihilation operators for these ghost excitations as
$a_Q({\bf k})$ and  $a_R({\bf k})$, and the creation operators which are
their adjoints in an indefinite metric space as $a_Q{}^\star({\bf k})$
and $a_R{}^\star({\bf k})$, respectively.  In order to produce ghost
states, these operators must commute with their respective adjoints.
Any state $a_Q{}^\star({\bf k})|p\rangle$, where $|p\rangle$ is a normed
Fock state consisting of electrons and observable photons, has the norm
$\langle p|a_Q({\bf k})a_Q{}^\star({\bf k})|p\rangle$.  This norm
vanishes when $a_Q({\bf k})$ and $a_Q{}^\star({\bf k})$ commute and
$a_Q({\bf k})$ annihilates the Fock space vacuum.  However, these ghost
operators cannot commute with {\it all\/} other operators, because then
they would become useless in representing the commutation relations of
the gauge fields.  A small generalization of the algebra of positive
metric Hilbert spaces suffices to satisfy all these requirements.  In
this generalized algebra, $a_Q({\bf k})$ and its adjoint
$a_Q{}^\star({\bf k})$ commute, as do $a_R({\bf k})$ and
$a_R{}^\star({\bf k})$.  In addition, we impose the commutation rules
\begin{equation} [a_Q({\bf k}), a_R{}^\star({\bf q})]=[a_R({\bf k}),
a_Q{}^\star({\bf q})] = \delta_{\bf kq}. \end{equation} The unit
operator in the one-particle ghost sector then is \begin{equation}
1_{\mbox{\rm\scriptsize OPG}} = \sum_{\bf k}[a_Q{}^\star({\bf
k})|0\rangle\langle 0|a_R({\bf k}) + a_R{}^\star({\bf
k})|0\rangle\langle 0|a_Q({\bf k})] \label{eq:1OPG} \end{equation} and,
in the $n$-particle ghost sector, the obvious generalization of
Eq.~(\ref{eq:1OPG}) applies. We represent {\bf A} as ${\bf A}={\bf
A}^{\mbox{\rm\scriptsize T}}+{\bf A}^{\mbox{\rm\scriptsize L}}$, i.e. as
a sum of a transverse and a longitudinal field.  The transverse field,
${\bf A}^{\mbox{\rm\scriptsize T}}$, is represented as a superposition
of propagating photons with the two helicity modes $\epsilon_i{}^n({\bf
k})$, where $i$ designates the spatial component and $n$ refers to one
of the two polarization modes.  The components of ${\bf
A}^{\mbox{\rm\scriptsize T}}$ are given as\footnote{$A_i$ designates the
spatial components of $A^\mu$ which, in relativistic notation, would be
represented as contravariant quantities.} \begin{equation}
A_i{}^{\mbox{\rm\scriptsize T}}({\bf x}) = \sum_{\bf
k}\frac{\epsilon_i{}^n({\bf k})}{\sqrt{2k}}\left[a_n({\bf k})e^{i{\bf
k\cdot x}} +a_n{}^\dagger({\bf k})e^{-i{\bf k\cdot x}}\right]
\label{eq:AiT} \end{equation} where $a_n{}^\dagger({\bf k})$ and
$a_n({\bf k})$ designate the creation and annihilation operators,
respectively, for transversely polarized photons with polarization mode
$n$ and momentum ${\bf k}$. The longitudinal gauge field, ${\bf
A}^{\mbox{\rm\scriptsize L}}$, is represented in terms of ghost
excitations as \begin{eqnarray} A_i{}^{\mbox{\rm\scriptsize L}}({\bf x})
&=& \sum_{\bf k}\frac{k_i}{2k^{3/2}}\left[a_R({\bf k})e^{i{\bf k\cdot
x}} +a_R{}^\star({\bf k})e^{-i{\bf k\cdot x}}\right]\nonumber\\ &+& (1-
\frac{\gamma}{2})\sum_{\bf k}\frac{k_i}{2k^{3/2}}\left[a_Q({\bf
k})e^{i{\bf k\cdot x}} +a_Q{}^\star({\bf k})e^{-i{\bf k\cdot x}}\right],
\label{eq:AiL} \end{eqnarray} and similarly, the time-like component is
represented as \begin{eqnarray} A_0({\bf x}) &=& \sum_{\bf
k}\frac{1}{2\sqrt{k}}\left[a_R({\bf k})e^{i{\bf k\cdot x}}
+a_R{}^\star({\bf k})e^{-i{\bf k\cdot x}}\right]\nonumber\\ &-& (1-
\frac{\gamma}{2})\sum_{\bf k}\frac{1}{2\sqrt{k}}\left[a_Q({\bf
k})e^{i{\bf k\cdot x}} +a_Q{}^\star({\bf k})e^{-i{\bf k\cdot x}}\right].
\label{eq:A0} \end{eqnarray} ${\bf \Pi}$, the momentum conjugate to {\bf
A} as well as the negative of the electric field, is represented as
\begin{eqnarray} \Pi_i({\bf x}) &=& -i\sum_{\bf k}\epsilon_i{}^n({\bf
k})\sqrt{\frac{k}{2}}\left[a_n({\bf k})e^{i{\bf k\cdot x}} -
a_n{}^\dagger({\bf k})e^{-i{\bf k\cdot x}}\right]\nonumber\\ &-&
i\sum_{\bf k}\frac{k_i}{\sqrt{k}}\left[a_Q({\bf k})e^{i{\bf k\cdot x}}
-a_Q{}^\star({\bf k})e^{-i{\bf k\cdot x}}\right]; \label{eq:Piexp}
\end{eqnarray} and the gauge-fixing field, $G({\bf x})$, is represented
as \begin{equation} G({\bf x}) = i\sum_{\bf k}\sqrt{k}\left[a_Q({\bf
k})e^{i{\bf k\cdot x}} -a_Q{}^\star({\bf k})e^{-i{\bf k\cdot x}}\right].
\label{eq:G} \end{equation} The standard representation of $\psi$ and
$\psi^\dagger$ in terms
of electron and positron creation and annihilation operators will be
implicitly assumed.  The choice of these representations of the gauge
fields is determined by a number of requirements.  They must implement
the equal-time commutation rules given in Eqs.~(\ref{eq:comAiPij}) and
(\ref{eq:comAoG}); and they must lead to an implementable Hilbert space.
The latter requirement is connected to the gauge-invariance of the
theory, and will be discussed later.

When the representations given in Eqs.~(\ref{eq:AiT})-(\ref{eq:G}) are
substituted into the Hamiltonians ${H}_0{}^{\mbox{\rm\scriptsize cov}}$
and ${H}_{\mbox{\rm\scriptsize I}}{}^{\mbox{\rm\scriptsize cov}}$ where
$H = \int d{\bf x}\ {\cal H}({\bf x})$, we obtain \begin{eqnarray}
{H}_0{}^{\mbox{\rm\scriptsize cov}} &=& \sum_{\bf
k}k\left[a_n{}^\dagger({\bf k})a_n({\bf k}) + a_R{}^\star({\bf
k})a_Q({\bf k})+a_Q{}^\star({\bf k})a_R({\bf k}) + \gamma
a_Q{}^\star({\bf k}) a_Q({\bf k})\right]\nonumber\\ &+&\sum_{\bf
q}\omega_q\left[e_{s}{}^\dagger({\bf q}) e_{s}({\bf q}) +
\bar{e}_{s}{}^\dagger({\bf q})\bar{e}_{s}({\bf q})\right]
\label{eq:H0cov} \end{eqnarray} where $\omega_q=\sqrt{q^2 + m^2}$ and
$m$ is the electron mass; $e_{s}({\bf q})$ and $e_{s}{}^\dagger({\bf
q})$ represent electron annihilation and creation operators,
respectively, and the barred symbols designate the corresponding
positron  operators.

We can construct a Fock space, $\{|h\rangle\}$, based on the
perturbative vacuum, $|0\rangle$, which is annihilated by all the
annihilation operators, $a_n({\bf k})$, $a_Q({\bf k})$ and $a_R({\bf
k})$, as well as, $e_s({\bf k})$ and $\bar{e}_s({\bf k})$. This
perturbative Fock space includes all multiparticle states, $|N\rangle$,
consisting of observable, propagating particles, i.e., electrons,
positrons, and photons, created when $e_s{}^\dagger({\bf k})$,
$\bar{e}_s{}^\dagger({\bf k})$, and $a_n{}^\dagger({\bf k})$,
respectively, act on $|0\rangle$. All such states, $|N\rangle$, are
eigenstates of $H_0{}^{\mbox{\rm\scriptsize cov}}$. States in which a
single variety of ghost creation operator acts on one of these
multiparticle states $|N\rangle$, such as $a_Q{}^\star({\bf
k})|N\rangle$ or $a_Q{}^\star({\bf k}_1)a_Q{}^\star({\bf
k}_2)|N\rangle$, have zero norm; they have no probability of being
observed, and have vanishing expectation values of energy, momentum, as
well as, of all other observables. We will designate the subspace of
$\{|h\rangle\}$ that consists of all states $|N\rangle$, and of all
states in which a chain of $a_Q{}^\star({\bf k})$ operators (but {\it
no\/} $a_R{}^\star({\bf k})$ operators) act on $|N\rangle$, as
$\{|n\rangle\}$. States in which both varieties of ghost appear
simultaneously, such as $a_Q{}^\star({\bf k}_1)a_R{}^\star({\bf
k}_2)|N\rangle$, also are in the Fock space $\{|h\rangle\}$; but because
these states have non-vanishing norms and contain ghosts, they are not
interpretable. Their appearance in the course of time evolution signals
a catastrophic defect in the theory.

The time evolution operator $\exp\left(-i{H}_0{}^{\mbox{\rm\scriptsize
cov}}t\right)$---which excludes the effect of the interaction
Hamiltonian---has the important property that, if it acts on a state
vector $|n_i\rangle$ in $\{|n\rangle\}$, it can only propagate it within
$\{|n\rangle\}$; but it can never translate it into the part of
$\{|h\rangle\}$ external to $\{|n\rangle\}$. We observe that the only
parts of ${H}_0{}^{\mbox{\rm\scriptsize cov}}$ that could possibly cause
a state vector to leave the subspace $\{|n\rangle\}$ are those that
contain either $a_R{}^\star({\bf k})$ or $a_R({\bf k})$ operators. The
only part of ${H}_0{}^{\mbox{\rm\scriptsize cov}}$ that has that feature
contains the combination of operators $\Gamma({\bf k}) =
a_R{}^\star({\bf k})a_Q({\bf k}) + a_Q{}^\star({\bf k})a_R({\bf k})$.
When $a_R({\bf k})$ acts on a state vector $|n_i\rangle$, it either
annihilates the vacuum or it annihilates one of the $a_Q{}^\star({\bf
k})$ operators in $|n_i\rangle$. In the latter case, $\Gamma$ replaces
the annihilated $a_Q{}^\star({\bf k})$ operator with an identical one.
When $a_Q({\bf k})$ acts on a state vector $|n_i\rangle$, it always
annihilates it. It is therefore impossible for $\Gamma$ to produce a
state vector external to $\{|n\rangle\}$, i.e. one in which an
$a_R{}^\star({\bf k})$ operator acts on $|n_i\rangle$. The only effect
of $\Gamma$ is to translate $|n_i\rangle$ states within $\{|n\rangle\}$.
Substitution of Eqs.~(\ref{eq:AiT})-(\ref{eq:G}) into
${H}_{\mbox{\rm\scriptsize I}}{}^{\mbox{\rm\scriptsize cov}}$ leads to
\begin{eqnarray} &&{H}_{\mbox{\rm\scriptsize I}}{}^{\mbox{\rm\scriptsize
cov}} = -\sum_{{\bf k}}\frac{1}{\sqrt{2k}}\left[a_n({\bf k})\,{\bf j}(-
{\bf k})\cdot\hat{\mbox{\boldmath $\epsilon$}}^n({\bf k}) +
a_n{}^\dagger({\bf k})\,{\bf j}({\bf k})\cdot\hat{\mbox{\boldmath
$\epsilon$}}^n({\bf k})\right]\nonumber\\ &-&(1-
\frac{\gamma}{2})\sum_{\bf k}\frac{1}{2\sqrt{k}}\left[a_Q({\bf
k})\left(j_0(-{\bf k}) + \frac{\bf k \cdot j(-k)}{|{\bf k}|}\right)+
a_Q{}^\star({\bf k})\left(j_0({\bf k}) + \frac{\bf k \cdot j(k)}{|{\bf
k}|}\right)\right]\nonumber\\ &+&\sum_{\bf
k}\frac{1}{2\sqrt{k}}\left[a_R({\bf k})\left(j_0(-{\bf k}) - \frac{\bf k
\cdot j(-k)}{|{\bf k}|}\right)+ a_R{}^\star({\bf k})\left(j_0({\bf k}) -
\frac{\bf k \cdot j(k)}{|{\bf k}|}\right)\right]. \label{eq:HIcov}
\end{eqnarray} In this expression of ${H}_{\mbox{\rm\scriptsize
I}}{}^{\mbox{\rm\scriptsize cov}}$, all gauge field excitations appear,
including creation and annihilation operators for {\it both\/} varieties
of ghosts. This notifies us that ${H}_{\mbox{\rm\scriptsize
I}}{}^{\mbox{\rm\scriptsize cov}}$ causes transitions from the ``safe''
subspace $\{|n\rangle\}$ into the part of the larger space occupied by
uninterpretable state vectors that nevertheless absorb probability
amplitude, energy and momentum. In
Sec.~\ref{sec:implementationofconstraints}, we will show how
implementation of the constraints prevents the catastrophic appearance
of these state vectors in the course of time evolution.

\section{The Perturbative Regime} \label{sec:perturbativeregime} The
perturbative theory involves the vertices dictated by the interaction
Hamiltonian given in Eq.~(\ref{eq:HIdensity}) and the propagators for
the interaction-picture operators $\psi(x)$, $\bar{\psi}(x)$, and
$A^\mu(x)$. The interaction-picture operators are given by $P(x) =
\exp\left(i{H}_0{}^{\mbox{\rm\scriptsize cov}}t\right)P({\bf
x})\exp\left(-i{H}_0{}^{\mbox{\rm\scriptsize cov}}t\right)$ and, in the
case of the gauge fields, they are given by \begin{eqnarray} A_i(x) &=&
\sum_{\bf k}\frac{\epsilon_i{}^n({\bf k})}{\sqrt{2k}}\left[a_n({\bf
k})e^{-ik_\mu x^\mu} +a_n{}^\dagger({\bf k})e^{ik_\mu
x^\mu}\right]\nonumber\\ &+& \sum_{\bf
k}\frac{k_i}{2k^{3/2}}\left[a_R({\bf k})e^{-ik_\mu x^\mu}
+a_R{}^\star({\bf k})e^{ik_\mu x^\mu}\right]\nonumber\\ &+& (1-
\frac{\gamma}{2})\sum_{\bf k}\frac{k_i}{2k^{3/2}}\left[a_Q({\bf k})e^{-
ik_\mu x^\mu} +a_Q{}^\star({\bf k})e^{ik_\mu x^\mu}\right]\nonumber\\
&-& \sum_{\bf k}\frac{i\gamma x_0 k_i}{2\sqrt{k}}\left[a_Q({\bf k})e^{-
ik_\mu x^\mu} -a_Q{}^\star({\bf k})e^{ik_\mu x^\mu}\right]
\label{eq:Ai(x)} \end{eqnarray} and \begin{eqnarray} A_0(x) &=&
\sum_{\bf k}\frac{1}{2\sqrt{k}}\left[a_R({\bf k})e^{-ik_\mu x^\mu}
+a_R{}^\star({\bf k})e^{ik_\mu x^\mu}\right]\nonumber\\ &-& (1-
\frac{\gamma}{2})\sum_{\bf k}\frac{1}{2\sqrt{k}}\left[a_Q({\bf k})e^{-
ik_\mu x^\mu} +a_Q{}^\star({\bf k})e^{ik_\mu x^\mu}\right]\nonumber\\
&-&\sum_{\bf k}\frac{i\gamma x_0 \sqrt{k}}{2}\left[a_Q({\bf k})e^{-
ik_\mu x^\mu} -a_Q{}^\star({\bf k})e^{ik_\mu x^\mu}\right].
\label{eq:A0(x)} \end{eqnarray} In Eqs.~(\ref{eq:Ai(x)}) and
(\ref{eq:A0(x)}), $k^\mu$ is ``on-shell,'' i.e., $k_0 = |{\bf k}|$. We
use $A_i(x)$ and $A_0(x)$ in the expression for the propagator,
\begin{equation} D^{\mu\nu}(x,y) = \langle 0|\mbox{\sf
T}(A^\mu(x)A^\nu(y))|0\rangle \end{equation} where $\mbox{\sf T}$
designates time-ordering, and where $|0\rangle$ is the perturbative
vacuum annihilated by all annihilation operators, $a_n({\bf k})$,
$a_Q({\bf k})$ and $a_R({\bf k})$, as well as, $e_s({\bf p})$ and
$\bar{e}_s({\bf p})$ for electrons and positrons, respectively. We
obtain the expression \cite{tomczak} \begin{eqnarray} D^{ij}(x,y) &=&
\int \frac{d{\bf k}}{(2\pi)^3}\left\{\frac{e^{-ik_\mu(x-
y)^\mu}}{2k}\left[\delta_{ij} - \frac{\gamma k_ik_j}{|{\bf
k}|^2}\left(1+ik(x_0-y_0)\right)\right]\Theta(x_0-y_0)\right.\nonumber\\
&+&\left. \frac{e^{ik_\mu(x-y)^\mu}}{2k}\left[\delta_{ij} - \frac{\gamma
k_ik_j}{|{\bf k}|^2}\left(1-ik(x_0-y_0)\right)\right]\Theta(y_0-
x_0)\right\}, \end{eqnarray} \begin{eqnarray} D^{00}(x,y) &=& -\int
\frac{d{\bf k}}{(2\pi)^3}\left\{\frac{e^{-ik_\mu(x-y)^\mu}}{2k}\left[1 -
\frac{\gamma}{2}\left(1-ik(x_0-y_0)\right)\right]\Theta(x_0-
y_0)\right.\nonumber\\ &-&\left.\frac{e^{ik_\mu(x-y)^\mu}}{2k}\left[1 -
\frac{\gamma}{2}\left(1+ik(x_0-y_0)\right)\right]\Theta(y_0-
x_0)\right\}, \end{eqnarray} and \begin{equation} D^{0i}(x,y) = -
i\gamma\int \frac{d{\bf k}}{(2\pi)^3}\ \frac{k_i(x_0-y_0)}{4|{\bf
k}|}\left[e^{-ik_\mu(x-y)^\mu}\Theta(x_0-y_0) - e^{ik_\mu(x-
y)^\mu}\Theta(y_0-x_0)\right], \end{equation} which can be represented
as \begin{equation} D^{\mu\nu}(x,y) = i\int \frac{d^4k}{(2\pi)^4}\
\frac{e^{-ik_\lambda(x-y)^\lambda}}{k_\lambda k^\lambda +
i\epsilon}\left[g^{\mu\nu}-\frac{\gamma k^\mu k^\nu}{k_\lambda k^\lambda
+ i\epsilon}\right]; \end{equation} $\gamma=0$ corresponds to the
Feynman gauge, $\gamma=1$ corresponds to the Landau gauge, and
$\gamma=-2$ corresponds to the Fried-Yennie gauge.

The propagator $D^{\mu\nu}(x,y)$, a corresponding one for the spinor
field, and the vertices $\gamma^\mu$, constitute the Feynman rules for
$S$-matrix elements in the covariant gauges. The representation of the
$S$-matrix in terms of Feynman rules is based on an expression for the
$S$-matrix given by \begin{equation} S_{f,i} = \delta_{fi} - 2\pi
i\delta(E_i - E_f)T_{f,i}, \end{equation} where the subscripts refer to
the initial and final states $|i\rangle$ and $|f\rangle$ respectively;
these states consist of creation operators for electrons, positrons, and
transversely polarized photons acting on the perturbative vacuum state
$|0\rangle$. $T_{f,i}$ can be represented as \begin{equation} T_{f,i} =
\sum_{n=1}^\infty T_{f,i}{}^n \end{equation} with the $n$th order
transition amplitude, $T_{f,i}{}^n$, given by \begin{equation}
T_{f,i}{}^n = \langle f|H_{\mbox{\rm\scriptsize
I}}{}^{\mbox{\rm\scriptsize cov}}\frac{1}{(E_i - H_{\mbox{\rm\scriptsize
0}}{}^{\mbox{\rm\scriptsize cov}})}H_{\mbox{\rm\scriptsize
I}}{}^{\mbox{\rm\scriptsize cov}}\frac{1}{(E_i - H_{\mbox{\rm\scriptsize
0}}{}^{\mbox{\rm\scriptsize cov}})}\cdots \frac{1}{(E_i -
H_{\mbox{\rm\scriptsize 0}}{}^{\mbox{\rm\scriptsize
cov}})}H_{\mbox{\rm\scriptsize I}}{}^{\mbox{\rm\scriptsize
cov}}|i\rangle. \label{eq:tfin} \end{equation} Standard procedures
transform Eq.~(\ref{eq:tfin}) into Feynman rules. \cite{example}

The essential point of this brief summary is that the states $|i\rangle$
and $|f\rangle$ {\it do not\/} implement Gauss's law except in the
physically uninteresting limit in which all interactions between charged
particles and the electromagnetic field have been eliminated.  It is
easily seen, for example, that in the case of the expectation value of
$\nabla \cdot {\bf E}$ in the perturbative one-electron state $|e_s({\bf
p})\rangle$, $\langle e_s({\bf p})|\partial_i\Pi_i({\bf x})|e_s({\bf
p})\rangle = 0$; $\langle e_s({\bf p})|\partial_i\Pi_i({\bf x}) +
j_0({\bf x})|e_s({\bf p})\rangle = 0$ would be required by Gauss's law.
In the next section, we will discuss measures that produce a Hilbert
space whose states implement Gauss's law for the complete theory---i.e.,
one that includes interactions between the electromagnetic field and
charged particles.

\section{Implementation of Constraints}
\label{sec:implementationofconstraints} We will define the operator
${\cal G}$, that expresses Gauss's law in the covariant gauge, as
\begin{equation} {\cal G}({\bf x}) = \partial_l\Pi_l({\bf x}) + j_0({\bf
x}) \label{eq:Gauss} \end{equation} so that Eq.~(\ref{eq:DoG}) can be
written as ${\cal G} = \partial_0G$. Substitution of
Eq.~(\ref{eq:Piexp}) into Eq.~(\ref{eq:Gauss}) leads to \begin{equation}
{\cal G}({\bf x}) = \sum_{\bf k}k^{3/2}[\Omega({\bf k})e^{i{\bf k\cdot
x}} + \Omega^\star({\bf k})e^{-i{\bf k\cdot x}}], \end{equation} where
$\Omega({\bf k})$ is defined as $\Omega({\bf k}) = a_Q({\bf k}) +
j_0({\bf k})/2k^{3/2}$. Similarly, we can express $\partial_\mu A^\mu$
as \begin{equation} \partial_\mu A^\mu({\bf x}) = i(1-\gamma)\sum_{\bf
k}\sqrt{k}\ [\Omega({\bf k})e^{i{\bf k\cdot x}} - \Omega^\star({\bf
k})e^{-i{\bf k\cdot x}}]. \end{equation} The pivotal fact about
$\Omega({\bf k})$ is that, since $[H,\Omega({\bf k})]=-|{\bf
k}|\Omega({\bf k})$, the corresponding Heisenberg operator, $\Omega({\bf
k},t)$, where \begin{equation} \Omega({\bf
k},t)=\exp\left(i{H}^{\mbox{\rm\scriptsize cov}}t\right)\Omega({\bf
k})\exp\left(-i{H}^{\mbox{\rm\scriptsize cov}}t\right), \end{equation}
has a $c$-number time-dependence given by $\Omega({\bf k},t) =
\Omega({\bf k})e^{-i|{\bf k}|t}$.

In general, the time-dependence of Heisenberg operators cannot be
represented by simple $c$-number expressions; it can only be expressed
by iterative expansions involving their interaction-picture equivalents.
$\Omega({\bf k},t)$, however, is an exception in this regard.  And it is
this property of $\Omega({\bf k})$ that makes it so useful in
implementing constraints.

We will now use $\Omega({\bf k})$ to define a subspace
$\{|\nu\rangle\}$, of another Fock space, in which all state vectors
$|\nu_i\rangle$ obey the condition \cite{gupta} \begin{equation}
\Omega({\bf k})|\nu\rangle = 0. \end{equation} We will call
$\{|\nu\rangle\}$ the ``physical subspace,'' because the constraints are
implemented in it. For all state vectors $|\nu\rangle$ and
$|\nu^\prime\rangle$ in this physical subspace $\{|\nu\rangle\}$,
$\langle\nu^\prime|{G}|\nu\rangle = 0$; similarly $\langle\nu^\prime
|\nabla G|\nu\rangle =0$ and  $\langle\nu^\prime |\partial_0
G|\nu\rangle = 0$. These equations demonstrate that, in the subspace
$\{|\nu\rangle\}$, both Gauss's law and the gauge condition hold and
Maxwell's equations are exactly satisfied. Moreover, a state vector
initially in the physical subspace will always remain entirely contained
in it, as it develops under time evolution. This follows from the $c$-
number time-dependence of the Heisenberg operator $\Omega({\bf k},t)$.
To complete the Fock space in which this physical subspace is embedded,
we note that there are unitary transformations, \cite{landovitz} $U =
e^{D_1}$ and $V = e^{D}$, for which \begin{equation} U^{-1}\Omega({\bf
k})U = V^{-1}\Omega({\bf k})V=a_Q({\bf k}), \end{equation} where
$D=D_1+D_2$ and where \begin{equation} D_1 = \sum_{\bf
k}\frac{1}{2k^{3/2}}\,[a_R({\bf k})j_0(-{\bf k})-a_R{}^\star({\bf
k})j_0({\bf k})] \end{equation} and \begin{equation} D_2 = \sum_{\bf
k}\frac{\phi({\bf k})}{2k^{3/2}}\,[a_Q({\bf k})j_0(-{\bf k})-
a_Q{}^\star({\bf k})j_0({\bf k})]; \end{equation} the form of $\phi({\bf
k})$ is arbitrary except that $\phi({\bf k}) = \phi(-{\bf k})$. Since
$\phi({\bf k})$ is an arbitrary function, and since with $\phi({\bf
k})=0$, $V$ reduces to $U$, we will not differentiate between $U$ and
$V$ hereafter, but regard $U$ as a special case of $V$.

We can use the unitary operator $V$ to construct the physical subspace
$\{|\nu\rangle\}$. We extract the previously defined subspace
$\{|n\rangle\}$ from the Hilbert space $\{|h\rangle\}$, and simply set
$|\nu_i\rangle = V|n_i\rangle$ for any operator $V$. Since $a_Q({\bf
k})|n_i\rangle=0$, $\Omega({\bf k})|\nu_i\rangle=0$ follows immediately
from the definition of $V$. As an alternate to explicitly constructing
the subspace $\{|\nu\rangle\}$, it is possible, and often most
convenient, to transpose the entire formalism into a unitarily
equivalent representation. In this unitarily equivalent representation,
we keep $\{|n\rangle\}$ as the physical subspace which, in the original,
untransformed representation, is given by $\{|\nu\rangle\}$. We must
then also transform all operators, so that for any operator $P$, $P
\rightarrow V^{-1}PV$. We will designate the transformed operators
$V^{-1}PV$ as $\tilde{P}$. In the transformed representation, $a_Q({\bf
k})|n_i\rangle=0$ is the equation that implements Gauss's law and the
gauge condition for the ``complete'' theory, i.e., for the theory of the
electromagnetic field interacting with the charged electron-positron
field.

In this unitarily equivalent representation, the theory gives rise to
the same equations of motion, and implements the same constraints as in
the original representation. The choice between the transformed and the
untransformed representations is entirely a matter of convenience. The
transformed Hamiltonian, $\tilde{H}^{\mbox{\rm\scriptsize cov}}$, is
given by \begin{equation} \tilde{H}^{\mbox{\rm\scriptsize cov}} =
H_{\mbox{\rm\scriptsize C}} + H_{\mbox{\rm\scriptsize Q}}
\label{eq:HcovHCHQ} \end{equation} where $H_{\mbox{\rm\scriptsize C}}$
and $H_{\mbox{\rm\scriptsize Q}}$ are given by \begin{eqnarray}
H_{\mbox{\rm\scriptsize C}} &=& \sum_{\bf k}ka_n{}^\dagger({\bf
k})a_n({\bf k})+\sum_{\bf q}\omega_q\left[e_{s}{}^\dagger({\bf q})
e_{s}({\bf q}) + \bar{e}_{s}{}^\dagger({\bf q})\bar{e}_{s}({\bf
q})\right]\nonumber\\ &+& \sum_{\bf k}\frac{j_0({\bf k})j_0(-{\bf
k})}{2k^2}-\sum_{\bf k}\frac{\epsilon_l{}^n({\bf
k})}{\sqrt{2k}}\left[a_n({\bf k})j_l(-{\bf k}) + a_n{}^\dagger({\bf
k})j_l({\bf k})\right] \end{eqnarray} and \begin{eqnarray}
H_{\mbox{\rm\scriptsize Q}} &=& \sum_{\bf k}k\left[a_R{}^\star({\bf
k})a_Q({\bf k})+a_Q{}^\star({\bf k})a_R({\bf k}) + \gamma
a_Q{}^\star({\bf k}) a_Q({\bf k})\right]\nonumber\\ &-&
(1+\frac{\gamma}{2})\sum_{\bf k}\frac{1}{2\sqrt{k}}\left[a_Q({\bf
k})j_0(-{\bf k}) + a_Q{}^\star({\bf k})j_0({\bf k})\right]\nonumber\\
&-&\sum_{\bf k}\frac{\phi({\bf k})}{2\sqrt{k}}\left[a_Q({\bf k})j_0(-
{\bf k}) + a_Q{}^\star({\bf k})j_0({\bf k})\right]\nonumber\\ &-&(1-
\frac{\gamma}{2})\sum_{\bf k}\frac{k_i}{2k^{3/2}}\left[a_Q({\bf
k})j_i(-{\bf k}) + a_Q{}^\star({\bf k})j_i({\bf k})\right]\nonumber\\
&+& \sum_{\bf k}\frac{\phi({\bf k})k_i}{2k^{3/2}}\left[a_Q({\bf
k})j_i(-{\bf k}) + a_Q{}^\star({\bf k})j_i({\bf k})\right],
\label{eq:HQ} \end{eqnarray} respectively. We can similarly transform
the operator-valued fields in this theory, leading to \begin{equation}
\tilde{A}_0({\bf x}) = A_0({\bf x}) + (1-\frac{\gamma}{2})\sum_{\bf
k}\frac{j_0({\bf k})e^{i{\bf k\cdot x}}}{2k^2} - \sum_{\bf
k}\frac{\phi({\bf k})j_0({\bf k})e^{i{\bf k\cdot x}}}{2k^2},
\label{eq:tildeA0} \end{equation} \begin{equation} \tilde{G}({\bf x}) =
G({\bf x}), \end{equation} \begin{equation} \tilde{A}_l({\bf x}) =
A_l({\bf x}), \label{eq:tildeAi} \end{equation} \begin{equation}
\tilde{\Pi}_l({\bf x}) = \Pi_l({\bf x}) + i\sum_{\bf k}\frac{k_lj_0({\bf
k})e^{i{\bf k\cdot x}}}{k^2}, \label{eq:Pitilde} \end{equation} and
\begin{equation} \tilde{\psi}({\bf x}) = {\psi}({\bf x})e^{{\cal D}({\bf
x})}, \label{eq:psitilde} \end{equation} with \begin{eqnarray} {\cal
D}({\bf x}) &=& -ie\int d{\bf y}\ \frac{1}{4\pi|{\bf x-
y}|}\left[\nabla\cdot{\bf A}({\bf y}) - \frac{1}{2}(1-
\frac{\gamma}{2})G({\bf y})\right]\nonumber\\ &+& \frac{ie}{2}\int d{\bf
y}\  G({\bf y})\frac{1}{\nabla^2}\phi({\bf x-y}) \end{eqnarray} where
\begin{equation} \phi({\bf y}) = \frac{1}{(2\pi)^3}\int
d{\mbox{\boldmath $\kappa$}}\ \phi(\mbox{\boldmath
$\kappa$})e^{i\mbox{\boldmath $\kappa$}\cdot{\bf y}}. \end{equation} The
currents therefore remain untransformed, with $\tilde{j}_0({\bf
x})=j_0({\bf x})$ and  $\tilde{\bf j}({\bf x})={\bf j}({\bf x})$

We make the following observations about
$\tilde{H}^{\mbox{\rm\scriptsize cov}}$. $H_{\mbox{\rm\scriptsize C}}$
is precisely the Hamiltonian for QED in the Coulomb gauge. Except for a
single piece proportional to $\Gamma({\bf k}) = a_R{}^\star({\bf
k})a_Q({\bf k}) + a_Q{}^\star({\bf k})a_R({\bf k})$,
$H_{\mbox{\rm\scriptsize Q}}$ consists entirely of Hermitian
combinations of $a_Q({\bf k})$ and $a_Q{}^\star({\bf k})$ operators, in
various combinations with each other and with the current densities
$j_0(\pm{\bf k})$ and ${\bf j}(\pm{\bf k})$. Every component of
$H_{\mbox{\rm\scriptsize Q}}$ is proportional to $a_Q({\bf k})$ or
$a_Q{}^\star({\bf k})$; and, most importantly, except for $\Gamma({\bf
k})$, there are no $a_R({\bf k})$ or $a_R{}^\star({\bf k})$ operators in
$H_{\mbox{\rm\scriptsize Q}}$. Previously we showed that, within the
subspace $\{|n\rangle\}$, $\Gamma({\bf k})$ can propagate
$a_Q{}^\star({\bf k})$ operators, but cannot cause any state vectors in
$\{|n\rangle\}$ to make transitions out of $\{|n\rangle\}$.
$H_{\mbox{\rm\scriptsize Q}}$ therefore cannot affect the observable
consequences of the time evolution imposed by the evolution operator
$\exp\left(-i\tilde{H}^{\mbox{\rm\scriptsize cov}}t\right)$.

Figure 1 is useful in illustrating the role that
$H_{\mbox{\rm\scriptsize Q}}$ can have in the time evolution of state
vectors. The Hilbert space $\{|h\rangle\}$ contains the subspace
$\{|n\rangle\}$, and $\{|n\rangle\}$ contains the set of states
$|N_i\rangle$, which constitute a quotient space consisting of
electrons, positrons and transversely polarized photons. The subspace
$\{|n\rangle\}$ includes all the $|N_i\rangle$ states, as well as all
other states in which chains of $a_Q{}^\star({\bf k})$ operators act on
$|N_i\rangle$ states. $\{|h\rangle\}$ contains all of $\{|n\rangle\}$,
as well as all other states in which chains of $a_R{}^\star({\bf k})$
and $a_Q{}^\star({\bf k})$ operators act on $\{|n\rangle\}$ states. When
the time evolution operator in the transformed representation,
$\exp\left(-i\tilde{H}^{\mbox{\rm\scriptsize cov}}t\right)$, acts on a
state $|N\rangle$, it time-translates it so that it moves on the sheet
representing the quotient space and also spreads along the fiber that
represents a set of states in $\{|n\rangle\}$. The time-translated state
extends along the fiber from a point on the sheet that represents the
quotient space; but it remains entirely within $\{|n\rangle\}$. The
states that are along the fiber, but no longer part of the quotient
space, all are zero-norm states. They never affect the trajectory of
states $\exp\left(-i\tilde{H}^{\mbox{\rm\scriptsize
cov}}t\right)|N\rangle$ within the quotient space because they have no
further interactions with the states that form the projection of
$\exp\left(-i\tilde{H}^{\mbox{\rm\scriptsize cov}}t\right)|N\rangle$
onto the quotient space. Because all the states in $\{|n\rangle\}$ that
are not part of the quotient space have zero norm, the theory is
manifestly unitary within the quotient space alone. And since the
interaction Hamiltonian in $\tilde{H}^{\mbox{\rm\scriptsize cov}}$ is
entirely devoid of $a_R{}^\star({\bf k})$ operators,
$H_{\mbox{\rm\scriptsize Q}}$ cannot be involved in the generation of
internal loops, and therefore also cannot change any of the radiative
corrections. In fact, if $H_{\mbox{\rm\scriptsize Q}}$  were entirely
eliminated from  $\tilde{H}^{\mbox{\rm\scriptsize cov}}$, there would be
no change in the trajectory of the point that represents the projection
of the state $\exp\left(-i\tilde{H}^{\mbox{\rm\scriptsize
cov}}t\right)|N\rangle$ onto the quotient space. All amplitudes $\langle
f|\exp\left(-i\tilde{H}^{\mbox{\rm\scriptsize cov}}t\right)|N\rangle$,
where $|f\rangle$ is one of the $|N_i\rangle$ states, are entirely
identical to the corresponding $\langle f|\exp\left(-
i{H}_{\mbox{\rm\scriptsize C}}t\right)|N\rangle$. Equation
(\ref{eq:Pitilde}) demonstrates that in the transformed representation,
$\langle n_b|(\nabla\cdot\tilde{\bf E}-\tilde{j}_0)|n_a\rangle =0$.
Gauss's law is implemented for the complete theory, which includes
interactions between the electromagnetic field and charged particles,
when constraints are imposed.

The condition $a_Q({\bf k})|n_i\rangle = 0$ and the subspace
$\{|n\rangle\}$ sometimes are also used in the formulation of the theory
in the untransformed representation. The perturbative formulation
summarized in Sec.~\ref{sec:perturbativeregime} is an illustration of
that practice. But use of the subspace $\{|n\rangle\}$ with the {\it
untransformed\/} Hamiltonian $H^{\mbox{\rm\scriptsize cov}}$, does not
guarantee the validity of Maxwell's equations at all times. In that
case, the constraint $a_Q({\bf k})|n_i\rangle = 0$ implements Gauss's
law {\it only in the interaction-free limit,\/} in which all
interactions between the electromagnetic field and the charged particles
have been eliminated. It is therefore natural to ask why the
perturbative theory outline in Sec.~III, and the Feynman rules to which
it gives rise, can be used to evaluate $S$-matrix elements. To answer
that question we compare the transition amplitude that leads to the
Feynman rules for covariant gauges, \begin{equation} T_{f,i} = \langle
f|H_{\mbox{\rm\scriptsize I}}{}^{\mbox{\rm\scriptsize cov}} +
H_{\mbox{\rm\scriptsize I}}{}^{\mbox{\rm\scriptsize cov}}(E_i -
H^{\mbox{\rm\scriptsize cov}} + i\epsilon)^{-1}H_{\mbox{\rm\scriptsize
I}}{}^{\mbox{\rm\scriptsize cov}}|i\rangle, \end{equation} with the
transition amplitude, $\tilde{T}_{f,i}$, that results when Gauss's law
is implemented. We can express $\tilde{T}_{f,i}$ in one of two forms. In
one form, we let $\tilde{H}^{\mbox{\rm\scriptsize cov}} =
H_0{}^{\mbox{\rm\scriptsize cov}} + \hat{H}_{\mbox{\rm\scriptsize
I}}{}^{\mbox{\rm\scriptsize cov}}$ so that
$(\tilde{H}^{\mbox{\rm\scriptsize cov}} - E_i)|i\rangle =
\hat{H}_{\mbox{\rm\scriptsize I}}{}^{\mbox{\rm\scriptsize
cov}}|i\rangle$, but $\hat{H}_{\mbox{\rm\scriptsize
I}}{}^{\mbox{\rm\scriptsize cov}} \neq V^{-1}H_{\mbox{\rm\scriptsize
I}}{}^{\mbox{\rm\scriptsize cov}}V$. We then find that \begin{equation}
\tilde{T}_{f,i} = \langle f|\hat{H}_{\mbox{\rm\scriptsize
I}}{}^{\mbox{\rm\scriptsize cov}} + \hat{H}_{\mbox{\rm\scriptsize
I}}{}^{\mbox{\rm\scriptsize cov}}(E_i - \tilde{H}^{\mbox{\rm\scriptsize
cov}} + i\epsilon)^{-1}\hat{H}_{\mbox{\rm\scriptsize
I}}{}^{\mbox{\rm\scriptsize cov}}|i\rangle. \end{equation}
Alternatively, we can express $H^{\mbox{\rm\scriptsize cov}}$ in the
form $H^{\mbox{\rm\scriptsize cov}} = {\mbox{\sf H}}_0 + {\mbox{\sf
H}}_{\mbox{\rm\scriptsize I}}$, such that the states in the subspace
$\{|\nu\rangle\}$ are eigenstates of $\mbox{\sf H}_0$, i.e., $(\mbox{\sf
H}_0-E_i)|\nu_i\rangle = 0$. Then \begin{equation} \tilde{T}_{f,i} =
\langle f|{\mbox{\sf H}}_{\mbox{\rm\scriptsize I}} + {\mbox{\sf
H}}_{\mbox{\rm\scriptsize I}}(E_i - {H}^{\mbox{\rm\scriptsize cov}} +
i\epsilon)^{-1}{\mbox{\sf H}}_{\mbox{\rm\scriptsize I}}|\nu_i\rangle.
\end{equation} We will show that, because the sets of states
$|\nu\rangle$ and $|n\rangle$ are unitarily equivalent, $T_{f,i}$ may
safely be substituted for the correct $\tilde{T}_{f,i}$ in the $S$-
matrix, although $T_{f,i}$ is not in general identical to
$\tilde{T}_{f,i}$. The argument proceeds as follows:
\cite{landovitz,haller1,sohn} We rewrite $T_{f,i}$ in the form
\begin{equation} T_{f,i} = \langle f|H_{\mbox{\rm\scriptsize
I}}{}^{\mbox{\rm\scriptsize cov}}|\Psi_i\rangle, \label{eq:Tfi}
\end{equation} where $|\Psi_i\rangle$ is the scattering state with
outgoing boundary conditions, \begin{equation} |\Psi_i\rangle = \left(1
+ (E_i - H^{\mbox{\rm\scriptsize cov}}+i\epsilon)^{-
1}H_{\mbox{\rm\scriptsize I}}{}^{\mbox{\rm\scriptsize
cov}}\right)|i\rangle. \end{equation} Similarly, we express
$\tilde{T}_{f,i}$ in the form \begin{equation} \tilde{T}_{f,i} = \langle
f|\hat{H}_{\mbox{\rm\scriptsize I}}{}^{\mbox{\rm\scriptsize
cov}}|\hat{\Psi}_i\rangle \label{eq:TildeTfi} \end{equation} where
$|\hat{\Psi}_i\rangle$ is the scattering state \begin{equation}
|\hat{\Psi}_i\rangle = \left(1 + (E_i - \tilde{H}^{\mbox{\rm\scriptsize
cov}}+i\epsilon)^{-1}\hat{H}_{\mbox{\rm\scriptsize
I}}{}^{\mbox{\rm\scriptsize cov}}\right)|i\rangle \end{equation} that
corresponds to the formulation in which Gauss's law has been
implemented.

It is easy to see that \begin{equation} \hat{H}_{\mbox{\rm\scriptsize
I}}{}^{\mbox{\rm\scriptsize cov}} = H_{\mbox{\rm\scriptsize
I}}{}^{\mbox{\rm\scriptsize cov}}V + (1-V)\hat{H}_{\mbox{\rm\scriptsize
I}}{}^{\mbox{\rm\scriptsize cov}} - H_0{}^{\mbox{\rm\scriptsize
cov}}(1-V) \end{equation} and \begin{equation} {H}_{\mbox{\rm\scriptsize
I}}{}^{\mbox{\rm\scriptsize cov}} = V\hat{H}_{\mbox{\rm\scriptsize
I}}{}^{\mbox{\rm\scriptsize cov}} + {H}_{\mbox{\rm\scriptsize
I}}{}^{\mbox{\rm\scriptsize cov}}(1-V) - (1-
V)H_0{}^{\mbox{\rm\scriptsize cov}}, \end{equation} and that, therefore,
\begin{equation} |\hat{\Psi}_i\rangle = V^{-1}|\Psi_i\rangle -
i\epsilon(E_i - \tilde{H}^{\mbox{\rm\scriptsize cov}} + i\epsilon)^{-
1}(V^{-1}-1)|i\rangle. \end{equation} Substitution of these relations
into Eqs.~(\ref{eq:Tfi}) and (\ref{eq:TildeTfi}) leads to
\begin{equation} T_{f,i} = \tilde{T}_{f,i} + (E_f - E_i){\cal
T}_{f,i}{}^{(\alpha)} + i\epsilon{\cal T}_{f,i}{}^{(\beta)}
\label{eq:tfialphabeta} \end{equation} where \begin{equation} {\cal
T}_{f,i}{}^{(\alpha)} = \langle f|(1-V)|\hat{\Psi}_i\rangle
\end{equation} and \begin{equation} {\cal T}_{f,i}{}^{(\beta)} = \langle
f |\left[H_{\mbox{\rm\scriptsize I}}{}^{\mbox{\rm\scriptsize cov}}
\frac{1}{E_i - H^{\mbox{\rm\scriptsize cov}} + i\epsilon}(1-V)-(1-
V)\frac{1}{E_i - \tilde{H}^{\mbox{\rm\scriptsize
cov}}+i\epsilon}\hat{H}_{\mbox{\rm\scriptsize
I}}{}^{\mbox{\rm\scriptsize cov}}\right]|i\rangle. \end{equation}
Equation (\ref{eq:tfialphabeta}) establishes that, although the
transition amplitudes $T_{f,i}$ and $\tilde{T}_{f,i}$ can differ, they
are the same ``on-shell'', when $E_i = E_f,$ provided that ${\cal
T}_{f,i}{}^{(\beta)}$ remains bounded as $\epsilon \rightarrow 0$. But
in the case of QED, we find that because of singularities in ${\cal
T}_{f,i}{}^{(\beta)}$, $i\epsilon{\cal T}_{f,i}{}^{(\beta)}$ generates a
series of additional $S$-matrix elements in which $D_1$ vertices connect
with $\hat{H}_{\mbox{\rm\scriptsize I}}{}^{\mbox{\rm\scriptsize cov}}$
or ${H}_{\mbox{\rm\scriptsize I}}{}^{\mbox{\rm\scriptsize cov}}$
vertices. Since neither $H_{\mbox{\rm\scriptsize
I}}{}^{\mbox{\rm\scriptsize cov}}$ nor $\hat{H}_{\mbox{\rm\scriptsize
I}}{}^{\mbox{\rm\scriptsize cov}}$ commute with
$H_0{}^{\mbox{\rm\scriptsize cov}}$, the propagator $(E_i-
H_0{}^{\mbox{\rm\scriptsize cov}}+i\epsilon)^{-1}$ cannot bypass
$H_{\mbox{\rm\scriptsize I}}{}^{\mbox{\rm\scriptsize cov}}$ and/or
$\hat{H}_{\mbox{\rm\scriptsize I}}{}^{\mbox{\rm\scriptsize cov}}$ to act
on $|i\rangle$ and/or $|f\rangle$ directly and reduce
Eq.~(\ref{eq:tfialphabeta}) to a trivial identity. If we set $\phi({\bf
k}) = 0$, we find that $D=D_1$, and since two $D_1$ vertices cannot
connect, the states $e^{D_1}|n\rangle$ are normalized states. We further
observe that the factor $(i\epsilon)^{-1}$ can arise only in self-energy
insertions to external electron lines. There they appear as new,
spurious self-energy insertions, in which propagators connect a $D_1$
with either an $\hat{H}_{\mbox{\rm\scriptsize
I}}{}^{\mbox{\rm\scriptsize cov}}$ or an ${H}_{\mbox{\rm\scriptsize
I}}{}^{\mbox{\rm\scriptsize cov}}$ vertex in a self-energy part that is
entirely disconnected from the rest of the $S$-matrix element. Multiple
$(i\epsilon)^{-n}$ contributions, with $n>1$, cannot arise when the
theory has been mass-renormalized, so that the energy continua of
$\hat{H}^{\mbox{\rm\scriptsize cov}}$ and ${H}^{\mbox{\rm\scriptsize
cov}}$ coincide. The effect of substituting $T_{f,i}$ for
$\tilde{T}_{f,i}$ therefore only produces extra, spurious, self-energy
insertions to external lines, and these are absorbed in the
renormalization constants. The physical predictions are not affected by
this substitution. These extra contributions from $i\epsilon{\cal
T}_{f,i}{}^{(\beta)}$ to the renormalization constants have long been
known and have been designated ``gauge-dependent parts'' of the
renormalization constants. \cite{johnson} In fact, when the constraints
are implemented, the renormalization constants are identical in the
different forms of the covariant gauge, as well as, in the Coulomb
gauge. The $i\epsilon{\cal T}_{f,i}{}^{(\beta)}$ contributions arise
because the constraints are not implemented in the perturbative theory
in covariant gauges, \cite{landovitz} and they are responsible for the
fact that the renormalized, rather than the unrenormalized $S$-matrix
elements are gauge-independent. \cite{bialynicki}

\section{QED in the Coulomb and the Covariant Gauges}

The preceding discussion provides us with a basis for understanding the
relation between QED in the covariant gauge and in the Coulomb gauge, as
well as among the different forms of the covariant gauge. Gauge
transformations can be implemented by the unitary operator ${\cal V} =
e^{\tau}$, where $\tau = i\int d{\bf x}\ {\cal G}({\bf x})\chi({\bf
x})$, in which $\chi$ is a $c$-number function. Under this gauge
transformation the gauge field and the charged particle field transform
as \begin{equation} A_\mu \rightarrow A_\mu - \partial_\mu\chi
\end{equation} and \begin{equation} \psi \rightarrow \psi e^{ie\chi}.
\end{equation} The unitary transform ${\cal V}$ can be used to study the
behavior of various operator-valued quantities under a $c$-number gauge
transformation within a particular gauge formulation. But it is more
complicated to understand the relation between two different gauge
formulations of QED. For example, it is not immediately obvious that
$H^{\mbox{\rm\scriptsize cov}}$, the Hamiltonian for the covariant
gauge, and the Coulomb gauge Hamiltonian, $H_{\mbox{\rm\scriptsize C}}$,
describe identical interactions among charged particles and photons. It
is known that $S$-matrix elements with external photons polarized in the
$k_\mu$ direction vanish, unless the regularization procedure used to
control divergences introduces non-vanishing contributions. These $S$-
matrix elements describe the creation of ``forbidden'' $R$-type photon
ghosts, and the condition that they vanish is often ascribed to
``gauge-invariance of the $S$-matrix.'' It is also known that in tree
approximation Feynman rules give identical results for $S$-matrix
elements in different gauges. When radiative corrections are included,
the renormalization program in non-covariant gauges is still highly
problematical, and the finite parts of divergent $S$-matrix elements are
still not well defined. \cite{leibbrandt} The gauge-invariance of $S$-
matrix elements and Feynman rules have been widely studied. But the
comparison of canonical formulations of QED in different gauges has not
been the subject of systematic investigation.

Equations~(\ref{eq:HcovHCHQ})--(\ref{eq:HQ}) demonstrate that QED in the
Coulomb and covariant gauges are identical theories, in the sense that
when $\tilde{H}^{\mbox{\rm\scriptsize cov}}$, the unitary transform of
$H^{\mbox{\rm\scriptsize cov}}$, is projected onto the quotient space
composed of state vectors $|N_i\rangle$, that projection is identical to
the Coulomb gauge Hamiltonian $H_{\mbox{\rm\scriptsize C}}$. There are
two differences between $H^{\mbox{\rm\scriptsize cov}}$ and
$H_{\mbox{\rm\scriptsize C}}$. The chief difference is that the particle
creation (and annihilation) operators that appear in the two
Hamiltonians refer to different excitations, though they are commonly
represented as though they were identical; and, when these creation
operators act on the vacuum, they represent different states in the
covariant and Coulomb gauges. In the covariant gauge formulation that
leads to the Hamiltonian $H^{\mbox{\rm\scriptsize cov}}$,
$e_s{}^\dagger({\bf q}){}|0\rangle$ represents a pure ``Dirac'' electron
totally devoid of all electric or magnetic fields. In the Coulomb gauge
formulation however, $e_s{}^\dagger({\bf q}){}|0\rangle$ represents an
electron accompanied by the longitudinal electric field ${\bf E}({\bf
x})= -\nabla\int d{\bf y}\ j_0({\bf y})/4\pi|{\bf x-y}|$. The unitary
transformation $P \rightarrow \tilde{P} = V^{-1}PV$ shifts to a
representation in which $e_s{}^\dagger({\bf q}){}|0\rangle$ represents
an electron accompanied by its Coulomb field, in the covariant gauge as
well. $\tilde{H}^{\mbox{\rm\scriptsize cov}}$ and
$H_{\mbox{\rm\scriptsize C}}$ therefore are expressed in terms of the
same particle excitation operators.

The remaining difference between $\tilde{H}^{\mbox{\rm\scriptsize cov}}$
and $H_{\mbox{\rm\scriptsize C}}$ is entirely in the ``ghost'' part of
the spectrum, i.e., along the fiber within the subspace $\{|n\rangle\}$
but not on the quotient space of states $|N_i\rangle$. The difference
between $\tilde{H}^{\mbox{\rm\scriptsize cov}}$ and
$H_{\mbox{\rm\scriptsize C}}$ affects the equations of motion of the
gauge field; but it has no effect on the time evolution of state vectors
in the quotient space, which alone contains state vectors that describe
physically observable configurations of particles.

That there is a difference in the renormalization program for QED in
different gauges is entirely consistent with the fact that, in the
perturbative theory, Gauss's law is not implemented. The differences in
the renormalization constants ensue from that fact. In non-covariant
gauges, these renormalization constants are frame-dependent, and this
contributes to the difficulties that beset the renormalization program
in these gauges. But QED in the covariant and the Coulomb gauges, as
quantum field theories, are entirely equivalent when Gauss's law is
implemented in both, as shown in the preceding section. Later, we will
extend this result to other gauges.

\section{QED in the spatial axial gauge} A gauge which presents an
interesting contrast to the covariant gauge is the spatial axial, or the
$A_3=0$ gauge. \cite{kummer,arnowittt} The photon propagator for this
gauge has been known for some time, but its canonical quantization has
not been discussed extensively. Since $A_0$ is not involved in the gauge
condition, the gauge fixing term cannot simultaneously serve the purpose
of imposing $A_3=0$ and providing $A_0$ with a canonically conjugate
momentum. Primary constraints on operator-valued fields therefore are
inevitable, unless the gauge is approached as a limit of a more general
axial gauge \cite{haller}. The Lagrangian for this model is\footnote{In
this gauge, in which we use non-relativistic notation for the gauge
field, $j_i$ and $A_i$ refer to the contravariant quantities and
$\partial_i$ to the covariant quantity.} \begin{equation} {\cal L} = -
\case 1/4 F_{ij}F_{ij} + \case 1/2 F_{0i}F_{0i} + j_iA_i - j_0A_0 - A_3G
+ \bar{\psi}(i\gamma_\mu\partial^\mu - m)\psi \end{equation} where
$F_{ij} = \partial_jA_i-\partial_iA_j$ and
$F_{0i}=\partial_0A_i+\partial_iA_0$. The Euler-Lagrange equations that
follow from this Lagrangian are \begin{equation} \partial_0F_{0i} -
\partial_jF_{ij} - j_i + \delta_{i3}G = 0, \end{equation}
\begin{equation} \partial_iF_{0i} + j_0 = 0 \end{equation} and
\begin{equation} A_3 = 0. \end{equation} The momenta conjugate to the
fields are given by $\Pi_i = \delta{\cal L}/\delta(\partial_0A_i) =
F_{0i}$, $\Pi_0 = \delta{\cal L}/\delta(\partial_0A_0) = 0$ and
$\Pi_{\mbox{\rm\scriptsize G}} = \delta{\cal L}/\delta(\partial_0G) =
0$. The charged particle field is treated precisely as in the covariant
gauge. $\Pi_0 \approx 0$ and $\Pi_{\mbox{\rm\scriptsize G}} \approx 0$
are primary constraints. We will use the sign $\approx$ to designate
weak equalities that hold only by virtue of constraints. Commutation
rules must be modified to be consistent with primary constraints, as
well as with further derived constraints. We use the Dirac-Bergmann
method to carry out this program. \cite{dirac,bergmann} We first form
the Hamiltonian density \begin{equation} {\cal H}^{\mbox{\rm\scriptsize
spat}} = \partial_0A_i \frac{\delta{\cal L}}{\delta(\partial_0A_i)} +
\partial_0\psi\frac{\delta{\cal L}}{\delta(\partial_0\psi)} - {\cal L} +
\Pi_0U_0 + \Pi_{\mbox{\rm\scriptsize G}}U_{\mbox{\rm\scriptsize G}},
\end{equation} which becomes \begin{eqnarray} {\cal
H}^{\mbox{\rm\scriptsize spat}} &=& \case 1/2 \Pi_i\Pi_i + \case 1/4
F_{ij}F_{ij} + A_0(\nabla\cdot{\bf \Pi} + j_0) + A_3G - {\bf j\cdot
A}\nonumber\\ &+& \psi^\dagger(-i\mbox{\boldmath $\alpha$}\cdot\nabla +
\beta m)\psi + \Pi_0U_0 + \Pi_{\mbox{\rm\scriptsize
G}}U_{\mbox{\rm\scriptsize G}} \label{eq:Hdenspat} \end{eqnarray} where
$U_0$ and $U_{\mbox{\rm\scriptsize G}}$ are undetermined $c$-number
fields. Secondary constraints are obtained by assuming canonical
(Poisson) commutation rules for $\Pi_0$ and $\Pi_{\mbox{\rm\scriptsize
G}}$, and by setting $i[H^{\mbox{\rm\scriptsize spat}}, \Pi_0] \approx
0$ and $i[H^{\mbox{\rm\scriptsize spat}}, \Pi_{\mbox{\rm\scriptsize G}}]
\approx 0$. The resulting secondary constraints are \begin{equation}
\nabla\cdot{\bf \Pi} + j_0 \approx 0, \end{equation} with ${\bf \Pi} =
-{\bf E}$, and \begin{equation} A_3 \approx 0, \end{equation}
respectively. Further, tertiary constraints, obtained from
$i[H^{\mbox{\rm\scriptsize spat}}, \nabla\cdot{\bf \Pi} + j_0] \approx
0$ and $i[H^{\mbox{\rm\scriptsize spat}}, A_3] \approx 0$, are
\begin{equation} \partial_3G \approx 0 \end{equation} and
\begin{equation} \partial_3A_0 - \Pi_3 \approx 0, \end{equation}
respectively. Since $i[H^{\mbox{\rm\scriptsize spat}}, \partial_3G]$ and
$i[H^{\mbox{\rm\scriptsize spat}},\Pi_3 - \partial_3A_0]$ involve the
$c$-number fields $U_0$ and $U_{\mbox{\rm\scriptsize G}}$, they do not
lead to any further constraints.

We define the constraint functionals $\xi_1 = \Pi_0$, $\xi_2 =
\nabla\cdot{\bf \Pi} + j_0$, $\xi_3 = \partial_3G$, $\xi_4 =
\Pi_{\mbox{\rm\scriptsize G}}$, $\xi_5 = A_3$, and $\xi_6 =
\partial_3A_0 - \Pi_3$. And we establish the matrix of ``Poisson''
commutators, in which each field is assumed to have canonical
commutation rules with its adjoint momentum, even though that commutator
may not be consistent with the constraints. We let ${\cal M}_{i,j}({\bf
x,y}) = [\xi_i({\bf x}),\xi_j({\bf y})]$, and obtain the matrix
\begin{equation} {\cal M}({\bf x,y}) = \left(\begin{array}{ccccccccccc}
0 & \ & 0 & \ & 0 & \ & 0 & \ & 0 & \ & i\frac{\partial}{\partial x_3}\\
0 & \ & 0 & \ & 0 & \ & 0 & \ & -i\frac{\partial}{\partial x_3} & \ &
0\\ 0 & \ & 0 & \ & 0 & \ & i\frac{\partial}{\partial x_3} & \ & 0 & \ &
0\\ 0 & \ & 0 & \ & -i\frac{\partial}{\partial x_3} & \ & 0 & \ & 0 & \
& 0\\ 0 & \ & -i\frac{\partial}{\partial x_3} & \ & 0 & \ & 0 & \ & 0 &
\ & -i\\ i\frac{\partial}{\partial x_3} & \ & 0 & \ & 0 & \ & 0 & \ & i
& \ & 0\end{array}\right)\delta({\bf x-y}). \end{equation} The six
$\xi_i$ constitute a second class system of constraints, so that the
matrix ${\cal M}({\bf x,y})$ has an inverse, which we define as ${\cal
Y}({\bf x,y})$; then $\int d{\bf z}\ {\cal M}_{i,n}({\bf x,z}){\cal
Y}_{n,j}({\bf z,y}) = \delta_{ij}\delta({\bf x-y})$. We find that
\begin{equation} {\cal Y}({\bf x,y})=\left(\begin{array}{ccccccccc} 0 &
-i\left(\frac{\partial}{\partial x_3}\right)^{-1} & 0 & \ \ & 0 & \ \ &
0 & \ \ & -i\\ i\left(\frac{\partial}{\partial x_3}\right)^{-1}  & 0 & 0
& \ \ & 0 & \ \ & i & \ \ & 0\\ 0  & 0 &  0 & \ \ & i & \ \ & 0 & \ \ &
0\\ 0 &  0  & -i &  \ \ & 0 & \ \ & 0 &  \ \ & 0\\ 0 &  i &   0 &  \ \ &
0 & \ \ & 0 &  \ \ & 0\\ -i &  0 &   0 &  \ \ & 0 &  \ \ & 0 &  \ \ &
0\end{array}\right)\left(\frac{\partial}{\partial x_3}\right)^{-
1}\delta({\bf x-y}). \end{equation} The constrained ``Dirac''
commutator, $[\chi_a({\bf x}), \chi_b({\bf y})]^{\mbox{\rm\scriptsize
D}}$, for two fields $\chi_a({\bf x})$ and $\chi_b({\bf y})$, is given
by \begin{equation} [\chi_a({\bf x}), \chi_b({\bf
y})]^{\mbox{\rm\scriptsize D}} = [\chi_a({\bf x}), \chi_b({\bf y})] -
\int d{\bf z}d{\bf z^\prime}\ [\chi_a({\bf x}), \xi_i({\bf z})]{\cal
Y}_{i,j}({\bf z,z^\prime})[\xi_j({\bf z^\prime}), \chi_b({\bf y})].
\end{equation} The Dirac commutators for QED in the $A_3 = 0$ gauge are
\begin{equation} [A_0({\bf x}),A_i({\bf y})]^{\mbox{\rm\scriptsize D}} =
i\left[\left(\frac{\partial}{\partial x_3}\right)^{-
2}\frac{\partial}{\partial x_i}\right]\delta({\bf x-y})\ \ \ \mbox{\rm
for $i = 1,2$}; \end{equation} \begin{equation} [A_0({\bf x}),\psi({\bf
y})]^{\mbox{\rm\scriptsize D}} = e\left[\left(\frac{\partial}{\partial
x_3}\right)^{-2}\psi({\bf y})\right]\delta({\bf x-y}); \end{equation}
\begin{equation} [A_i({\bf x}),\Pi_j({\bf y})]^{\mbox{\rm\scriptsize D}}
= i\left[\delta_{ij}-\delta_{j3}\left(\frac{\partial}{\partial
x_3}\right)^{-1}\frac{\partial}{\partial x_i}\right]\delta({\bf x-y});
\label{eq:AiPijD} \end{equation} and \begin{equation} [\Pi_3({\bf
x}),\psi({\bf y})]^{\mbox{\rm\scriptsize D}} =
e\left[\left(\frac{\partial}{\partial x_3}\right)^{-1}\psi({\bf
y})\right]\delta({\bf x-y}). \end{equation} The Dirac commutators
$[A_3({\bf x}), \varphi({\bf y})]^{\mbox{\rm\scriptsize D}} = 0$ for any
$\varphi({\bf y})$; in fact, $[\xi_i({\bf x}), \varphi({\bf
y})]^{\mbox{\rm\scriptsize D}} = 0$ for any $\varphi({\bf y})$ and any
of the six $\xi_i({\bf x})$.

These Dirac commutators demonstrate that relationships exist among the
constrained quantities, which can be exploited to simplify the
Hamiltonian by reducing the number of independent quantities when
constraints are imposed. In the case of the $A_3 = 0$ gauge, these
relations are \begin{equation} A_0({\bf x}) \approx -
\left(\frac{\partial}{\partial x_3}\right)^{-
2}\left[\frac{\partial}{\partial x_n}\Pi_n({\bf x}) + j_0({\bf
x})\right], \label{eq:Aosimeq1} \end{equation} where the summation
extends over $n=1$ and 2 only, and \begin{equation} A_0({\bf x}) \approx
\left(\frac{\partial}{\partial x_3}\right)^{-1}\Pi_3({\bf x}).
\label{eq:Aosimeq2} \end{equation} These same equalities also follow
simply, on the classical level, from setting $A_3 = 0$ in Maxwell's
equations. When these relations are substituted in
Eq.~(\ref{eq:Hdenspat}), we obtain \begin{eqnarray} {\cal
H}^{\mbox{\rm\scriptsize spat}} &\approx& \case 1/2 \Pi_i\Pi_i + \case
1/4 F_{ij}F_{ij} - j_iA_i + \psi^\dagger(-i\mbox{\boldmath
$\alpha$}\cdot\nabla + \beta m)\psi\nonumber\\ &-& \frac{1}{2}
\left[\left(\frac{\partial}{\partial x_i}\Pi_i({\bf x}) + j_0({\bf
x})\right)\left(\frac{\partial}{\partial x_3}\right)^{-
2}\left(\frac{\partial}{\partial x_j}\Pi_j({\bf x})+j_0({\bf
x})\right)\right]\nonumber\\ &+&
\frac{1}{2}\left(\frac{\partial}{\partial x_3}A_i({\bf
x})\right)\left(\frac{\partial}{\partial x_3}A_i({\bf x})\right),
\label{eq:Hdensityspat} \end{eqnarray} where the summation now extends
only over $i, j = 1$ and 2, and where an implicit integration by parts
has been included. When we interpret the time derivative $\partial_0$ as
the commutator $\partial_0 = i[H^{\mbox{\rm\scriptsize spat}}, \ ]$, we
find that the Hamiltonian $H^{\mbox{\rm\scriptsize spat}} = \int d{\bf
x}\ {\cal H}^{\mbox{\rm\scriptsize spat}}({\bf x})$ reproduces Maxwell's
equations with the constraints imposed. Thus we obtain \begin{equation}
\partial_0\Pi_i - \partial_jF_{ij} - \partial_3\partial_3A_i - j_i = 0
\end{equation} and \begin{equation} \partial_0\Pi_3 -
\partial_3\partial_iA_i -j_3 = 0 \end{equation} for $i = 1,2$.

We note that the Hamiltonian $H^{\mbox{\rm\scriptsize spat}}$ is not
rotationally invariant, and that it bears very little resemblance to the
Hamiltonians $H^{\mbox{\rm\scriptsize cov}}$ or $H_{\mbox{\rm\scriptsize
C}}$, the covariant gauge and Coulomb gauge Hamiltonians respectively.
It is therefore relevant to ask to what extent $H^{\mbox{\rm\scriptsize
spat}}$, the Hamiltonian for QED in the $A_3=0$ gauge, describes the
same physics as $H^{\mbox{\rm\scriptsize cov}}$ or
$H_{\mbox{\rm\scriptsize C}}$.

To investigate this question, we expand the gauge field and the
canonical momenta in terms of photon creation and annihilation
operators. We must choose a representation that is consistent with the
constraint $A_3 = 0$, as well as with the canonical expression for the
magnetic field, \begin{equation} {\bf B}({\bf x}) = \sum_{\bf
k}\frac{i{{\bf k} \times \mbox{\boldmath $\epsilon$}^n({\bf
k})}}{\sqrt{2k}}\left[a_n({\bf k})e^{i{\bf k \cdot x}} -
a_n{}^\dagger({\bf k})e^{-i{\bf k \cdot x}}\right]. \end{equation} Since
${\bf B}({\bf x})$ is gauge invariant, it should have the identical form
in terms of photon creation and annihilation operators as in every other
gauge. A representation of $A_i({\bf x})$ that satisfies these
conditions is \begin{equation} A_i({\bf x}) = \sum_{\bf
k}\frac{1}{\sqrt{2k}}\left[\epsilon_i{}^n({\bf k}) -
\frac{k_i}{k_3}\epsilon_3{}^n({\bf k})\right]\left[a_n({\bf k})e^{i{\bf
k \cdot x}} + a_n{}^\dagger({\bf k})e^{-i{\bf k \cdot x}}\right].
\label{eq:Aisimeq} \end{equation} The representation for the canonically
conjugate momentum is again constrained by the gauge-invariance of the
electric field and by Gauss's law. We therefore represent $\Pi_i$, for
$i=1$ and 2, as in all other gauges, by \begin{equation} \Pi_i({\bf x})
= -i\sum_{\bf k}\epsilon_i{}^n({\bf k})\sqrt{\frac{k}{2}}\left[a_n({\bf
k})e^{i{\bf k \cdot x}} - a_n{}^\dagger({\bf k})e^{-i{\bf k \cdot
x}}\right]. \label{eq:Pisimeq} \end{equation} For $i=3$, we use
Eqs.~(\ref{eq:Aosimeq1}) and (\ref{eq:Aosimeq2}) and obtain
\begin{equation} \Pi_3({\bf x}) = -i\sum_{\bf k}\epsilon_3{}^n({\bf
k})\sqrt{\frac{k}{2}}\left[a_n({\bf k})e^{i{\bf k \cdot x}} -
a_n{}^\dagger({\bf k})e^{-i{\bf k \cdot x}}\right]-\partial_3{}^{-
1}j_0({\bf x}). \end{equation} These representations are manifestly
consistent with Gauss's law. When we use Eqs.~(\ref{eq:Aisimeq}) and
(\ref{eq:Pisimeq}) to evaluate the commutator $[A_i({\bf x}),\Pi_j({\bf
y})]$, the result reproduces the Dirac commutator given in
Eq.~(\ref{eq:AiPijD}), without requiring any ghost components for either
$A_i({\bf x})$ or $\Pi_j({\bf y})$.  When we substitute
Eqs.~(\ref{eq:Aisimeq}) and (\ref{eq:Pisimeq}) into
(\ref{eq:Hdensityspat}), we obtain \begin{equation}
H^{\mbox{\rm\scriptsize spat}} = H_0{}^{\mbox{\rm\scriptsize spat}} +
H_{\mbox{\rm\scriptsize I}}{}^{\mbox{\rm\scriptsize spat}}
\end{equation} where \begin{equation} H_0{}^{\mbox{\rm\scriptsize spat}}
= \sum_{n} ka_n{}^\dagger({\bf k})a_n({\bf k}) + \int d{\bf x}\
\psi^\dagger({\bf x})(\beta m - i\mbox{\boldmath
$\alpha$}\cdot\nabla)\psi({\bf x}), \end{equation} and the summation
over $n$ extends over the two photon helicity modes of the photon; and
\begin{equation} H_{\mbox{\rm\scriptsize I}}{}^{\mbox{\rm\scriptsize
spat}} = -\int d{\bf x}\ \left[j_i({\bf x})A_i({\bf x}) +
\partial_i\Pi_i({\bf x})\left(\frac{\partial}{\partial x_3}\right)^{-
2}j_0({\bf x}) + \frac{1}{2}j_0({\bf x})\left(\frac{\partial}{\partial
x_3}\right)^{-2}j_0({\bf x})\right], \end{equation} where the summation
extends over $i=1$ and 2 only.

We observe that $H_0{}^{\mbox{\rm\scriptsize spat}}$ represents the
kinetic energies of the non-interacting photons and electrons correctly;
but the interactions represented by $H_i{}^{\mbox{\rm\scriptsize spat}}$
are still frame-dependent, lack rotational invariance, and are not
manifestly equivalent to the covariant and Coulomb gauges. We can
demonstrate that these features stem from the fact that, in this $A_3=0$
gauge formulation, the state $e_s{}^\dagger({\bf q}){}|0\rangle$
represents an electron with a physically unrealistic, severely
asymmetric electric field. For example, we note that the expectation
value of the electric field for the charged particle state
$e_s{}^\dagger({\bf q}){}|0\rangle$ is \begin{equation} -\langle
e_s({\bf p})|\Pi_i({\bf x})|e_s({\bf p})\rangle = 0 \end{equation} for
$i=1,2$ and \begin{equation} -\langle e_s({\bf p})|\Pi_3({\bf
x})|e_s({\bf p})\rangle = -\langle e_s({\bf
p})|\left(\frac{\partial}{\partial x_3}\right)^{-1}j_0({\bf x})|e_s({\bf
p})\rangle. \end{equation} Although formally Gauss's law is preserved by
these equations, the spatial asymmetry of the electric field indicates
that the Hilbert space $\{|n\rangle\}$ is not appropriate for this
representation of the charged fermion field. We therefore need to
construct a Hilbert space that satisfies the requirement of spatial
symmetry as well as the implementation of constraints. We can construct
such a Hilbert space by transforming the states in the subspace
$\{|n\rangle\}$ to establish the states \begin{equation} |\phi_i\rangle
= e^{-\Delta}|n_i\rangle \label{eq:phii} \end{equation} with $\Delta$
given by \begin{equation} \Delta = \sum_{\bf k}
\frac{\epsilon_3{}^n({\bf k})}{k_3\sqrt{2k}}\left[a_n({\bf k})j_0(-{\bf
k}) - a_n{}^\dagger({\bf k})j_0({\bf k})\right]. \end{equation} The
states $|\phi_i\rangle$ incorporate spatial asymmetries that just
compensate for those in $H_{\mbox{\rm\scriptsize
I}}{}^{\mbox{\rm\scriptsize spat}}$. As in the case of the covariant
gauge, we can choose to apply the unitary transform to the operators
instead of to the states, and generate the transformed operators $P
\rightarrow \bar{P}=e^\Delta P e^{-\Delta}$. The expressions for the
transformed gauge fields are \begin{eqnarray} \bar{A}_i({\bf x}) =
\sum_{\bf k}\frac{1}{\sqrt{2k}}\left[\epsilon_i{}^n({\bf k}) -
\frac{k_i}{k_3}\epsilon_3{}^n({\bf k})\right]\left[a_n({\bf k})e^{i{\bf
k \cdot x}} + a_n{}^\dagger({\bf k})e^{-i{\bf k \cdot x}}\right],
\label{eq:Aihat} \end{eqnarray} \begin{equation} \bar{A}_0({\bf x}) = -
\sum_{\bf k}\frac{\epsilon_3{}^n({\bf
k})}{k_3}\sqrt{\frac{k}{2}}\left[a_n({\bf k})e^{i{\bf k\cdot x}} -
a_n{}^\dagger({\bf k})e^{-i{\bf k\cdot x}}\right] -
\frac{1}{\nabla^2}\,j_0({\bf x}) \label{eq:A0hat} \end{equation} where
$i=1,2$. The correspondingly transformed electric field obtained from
${\bf \Pi}=-{\bf E}$ and from \begin{equation} \bar{\Pi}_i({\bf x}) =
\Pi_i({\bf x}) - \partial_i\nabla^{-2}j_0({\bf x}) \end{equation} for
$i=1,2$, and \begin{equation} \bar{\Pi}_3({\bf x}) = -i\sum_{\bf
k}\epsilon_3{}^n({\bf k})\sqrt{\frac{k}{2}}\left[a_n({\bf k})e^{i{\bf
k\cdot x}} - a_n{}^\dagger({\bf k})e^{-i{\bf k\cdot x}}\right] -
\frac{\partial_3}{\nabla^2}\,j_0({\bf x}) \end{equation} shows that for
the transformed representation, spatial symmetry is restored and Gauss's
law is implemented in the subspace $\{|n\rangle\}$. The transformed
Hamiltonian $\bar{H}^{\mbox{\rm\scriptsize spat}}$ takes the form
\begin{equation} \bar{H}^{\mbox{\rm\scriptsize spat}} =
H_0{}^{\mbox{\rm\scriptsize spat}} - \sum_{{\bf k},
i}\frac{\epsilon_i{}^n({\bf k})}{\sqrt{2k}}\left[a_n({\bf k})j_i(-{\bf
k}) + a_n{}^\dagger({\bf k})j_i({\bf k})\right]+ \int d{\bf x}d{\bf y}\
\frac{j_0({\bf x})j_0({\bf y})}{8\pi|{\bf x-y}|} \end{equation} where
the $\sum_i$ here extends from $i=1$ to 3.
$\bar{H}^{\mbox{\rm\scriptsize spat}}$ is rotationally symmetric and
manifestly identical to $H_{\mbox{\rm\scriptsize C}}$, the Hamiltonian
for the Coulomb gauge.

The propagator for the perturbative theory is evaluated from the vacuum
state $|0\rangle$ that is part of the subspace $\{|n\rangle\}$ and from
the interaction-picture gauge field operators, \begin{equation} A_i(x) =
\sum_{\bf k}\frac{1}{\sqrt{2k}}\left(\epsilon_i{}^n({\bf k}) -
\frac{k_i}{k_3}\epsilon_3{}^n({\bf k})\right)\left[a_n({\bf k})e^{i({\bf
k\cdot x}-|{\bf k}|t)} +a_n{}^\dagger({\bf k})e^{-i({\bf k\cdot x}-|{\bf
k}|t)}\right] \end{equation} and \begin{equation} A_0(x) \approx -
\left(\frac{\partial}{\partial x_3}\right)^{-2}\frac{\partial}{\partial
x_i}\Pi_i(x) \end{equation} where the summation extends over $i=1$ and
2. The resulting propagators are $D_{ij}(x,y)=\langle 0|{\sf
T}\left(A_i(x)A_j(y)\right)|0\rangle$, given by \begin{equation}
D_{ij}(x,y)=\sum_{\bf k}\frac{1}{2k}\left(\delta_{ij} +
\frac{k_ik_j}{k_3{}^2}\right)e^{i\left[{\bf k\cdot(x-y)}-k|x_0-
y_0|\right]}, \end{equation} for $i=1$ or 2, and $D_{ij}(x,y)=0$ when
either $i$ or $j$ is in the $z$-direction. The propagator
$D_{i0}(x,y)=\langle 0|{\sf T}\left(A_i(x)A_0(y)\right)|0\rangle$ for
$i=1$ or $2$ is \begin{equation} D_{i0}(x,y) = \sum_{\bf
k}\frac{k_i}{2k_3{}^2}e^{i{\bf k\cdot(x-y)}}\left[e^{-ik(x_0-
y_0)}\Theta(x_0-y_0) - e^{ik(x_0-y_0)}\Theta(y_0-x_0)\right];
\end{equation} similarly, $D_{00}(x,y)$ is given by \begin{equation}
D_{00}(x,y) = \sum_{\bf k}\frac{k}{2k_3{}^2}\left(1-
\frac{k_3{}^2}{k^2}\right)e^{i\left[{\bf k\cdot(x-y)}-k|x_0-
y_0|\right]}. \end{equation} It is straightforward to show that these
expressions for the propagator can be obtained from \begin{equation}
D_{\mu\nu}(x,y) = \frac{1}{(2\pi)^4}\int d^4k\ D_{\mu\nu}(k)e^{-
ik_\lambda(x-y)^\lambda}, \end{equation} and from the axial gauge
propagator, \begin{equation} D_{\mu\nu}(k) = \frac{-i}{k_\lambda
k^\lambda + i\epsilon}\left(g_{\mu\nu} - \frac{k_\mu n_\nu + k_\nu
n_\mu}{k_\lambda n^\lambda} + \frac{n_\lambda n^\lambda k_\mu
k_\nu}{(k_\lambda n^\lambda)^2}\right), \end{equation} where $n_\mu =
\delta_{\mu 3}$.

In the spatial axial gauge, the perturbative Feynman rules are obtained
by combining the Hamiltonian $H^{\mbox{\rm\scriptsize spat}}$ with the
states $|n_i\rangle$, instead of the states $|\phi_i\rangle$ needed for
equivalence with the rotationally invariant theory in the Coulomb gauge.
However, in this case as in the covariant gauges, the argument presented
in Sec.~\ref{sec:implementationofconstraints} maintains the validity of
the $S$-matrix when the $|n_i\rangle$ are substituted for the
corresponding $|\phi_i\rangle$, except in so far as the renormalization
of QED in the spatial axial gauge is affected by this substitution. It
is also worth noting that, since \begin{equation} \langle
0|\Delta^2|0\rangle = \frac{1}{(2\pi)^3}\int d{\bf k}\
\frac{\epsilon_3{}^n({\bf k})\epsilon_3{}^n({\bf
k})}{2kk_3{}^2}\,j_0({\bf k})j_0(-{\bf k}), \end{equation} $\langle
\phi_i|\phi_i\rangle$ will not, in general, be finite; $|\phi_i\rangle$
therefore cannot be assumed to be a normalizable state.

\section{Quantum Gauge Transformations} The mathematical apparatus we
have developed in the preceding sections of this work now enables us to
implement gauge transformations by adding $A_\mu$ in one gauge to the
four-dimensional gradient of an operator-valued field, in order to
arrive at $A^\prime_\mu$, the gauge field in another gauge. The new
gauge, $A^\prime_\mu$, may differ from its usual version only by being
embedded in the Hilbert space of the original gauge $A_\mu$, which may,
for example, be the space $\{|n\rangle\}$ instead of the quotient space
of states $|N_i\rangle$. Thus $A^\prime_\mu$ may have a spurious
component whose operator content will be limited to $a_Q({\bf k})$ and
$a_Q{}^\star({\bf k})$ excitation operators. We are able to carry out
such operator gauge transformations, because the Hamiltonians, as well
as all operator-valued fields in the gauges being considered, have been
brought into what we will call ``common form.'' In this common form,
when excitation operators such as $e_{s}{}^\dagger({\bf p})$ and
$e_{s}({\bf p})$ act on the perturbative vacuum or on the states built
on it, they create and annihilate particle modes with identical
properties in all gauges. In common form, the Hamiltonians for the
Coulomb, covariant and spatial axial gauges are $H_{\mbox{\rm\scriptsize
C}}$, $\tilde{H}^{\mbox{\rm\scriptsize cov}}$, and
$\bar{H}^{\mbox{\rm\scriptsize spat}}$, respectively (with
$\bar{H}^{\mbox{\rm\scriptsize spat}}=H_{\mbox{\rm\scriptsize C}}$); and
all three of these Hamiltonians generate the identical time displacement
for a state vector $|N_i\rangle$ within the previously defined quotient
space.

We define the operator-valued field $\chi^{\mbox{\rm\scriptsize cov
$\rightarrow$ C}}$ as \begin{eqnarray} \chi^{\mbox{\rm\scriptsize cov
$\rightarrow$ C}} &=& i\sum_{\bf k}\frac{1}{2k^{3/2}}\left[a_R({\bf
k})e^{i{\bf k\cdot x}}-a_R{}^\star({\bf k})e^{-i{\bf k\cdot
x}}\right]\nonumber\\ &+& i(1-\frac{\gamma}{2})\sum_{\bf
k}\frac{1}{2k^{3/2}}\left[a_Q({\bf k})e^{i{\bf k\cdot x}}-
a_Q{}^\star({\bf k})e^{-i{\bf k\cdot x}}\right] \end{eqnarray} and note
that its gradient is given by \begin{eqnarray}
\partial_i\chi^{\mbox{\rm\scriptsize cov $\rightarrow$ C}} &=& -
\sum_{\bf k}\frac{k_i}{2k^{3/2}}\left[a_R({\bf k})e^{i{\bf k\cdot
x}}+a_R{}^\star({\bf k})e^{-i{\bf k\cdot x}}\right]\nonumber\\ &-& (1-
\frac{\gamma}{2})\sum_{\bf k}\frac{k_i}{2k^{3/2}}\left[a_Q({\bf
k})e^{i{\bf k\cdot x}}+a_Q{}^\star({\bf k})e^{-i{\bf k\cdot x}}\right]
\end{eqnarray} and $\partial_0\chi^{\mbox{\rm\scriptsize cov
$\rightarrow$ C}}=i[\tilde{H}^{\mbox{\rm\scriptsize
cov}},\chi^{\mbox{\rm\scriptsize cov $\rightarrow$ C}}]$, so that
\begin{eqnarray} \partial_0\chi^{\mbox{\rm\scriptsize cov $\rightarrow$
C}} &=& -(1+\frac{\gamma}{2})\sum_{\bf k}\frac{1}{2k^2}j_0({\bf
k})e^{i{\bf k\cdot x}}-\sum_{\bf k}\frac{\phi({\bf k})}{2k^2}j_0({\bf
k})e^{i{\bf k\cdot x}}\nonumber\\ &+&\sum_{\bf
k}\frac{1}{2\sqrt{k}}\left[a_R({\bf k})e^{i{\bf k\cdot
x}}+a_R{}^\star({\bf k})e^{-i{\bf k\cdot x}}\right]\nonumber\\ &+&
(1+\frac{\gamma}{2})\sum_{\bf k}\frac{1}{2\sqrt{k}}\left[a_Q({\bf
k})e^{i{\bf k\cdot x}}+a_Q{}^\star({\bf k})e^{-i{\bf k\cdot x}}\right].
\end{eqnarray} It follows from these expressions for
$\partial_\mu\chi^{\mbox{\rm\scriptsize cov $\rightarrow$ C}}$, that for
$\tilde{A}_\mu{}^{\mbox{\rm\scriptsize cov}}({\bf x})$, given in
Eqs.~(\ref{eq:AiT}), (\ref{eq:AiL}), (\ref{eq:A0}), (\ref{eq:tildeA0})
and (\ref{eq:tildeAi}), \begin{equation}
\tilde{A}_\mu{}^{\mbox{\rm\scriptsize cov}}({\bf x}) -
\partial_\mu\chi^{\mbox{\rm\scriptsize cov $\rightarrow$ C}} =
{A}_\mu{}^{\mbox{\rm\scriptsize C}}({\bf x}) + {\cal
A}_\mu{}^{\mbox{\rm\scriptsize C}}({\bf x}). \end{equation}
${A}_\mu{}^{\mbox{\rm\scriptsize C}}({\bf x})$ is the form that
${A}_\mu({\bf x})$ takes in the Coulomb gauge.
${A}_i{}^{\mbox{\rm\scriptsize C}}({\bf x}) =
{A}_i{}^{\mbox{\rm\scriptsize T}}({\bf x})$, the transverse part of
$A_i({\bf x})$; ${A}_0{}^{\mbox{\rm\scriptsize C}}({\bf x})$ is given by
\begin{equation} {A}_0{}^{\mbox{\rm\scriptsize C}}({\bf x}) = \int d{\bf
y}\ \frac{j_0({\bf y})}{4\pi|{\bf x-y}|}. \end{equation} ${\cal
A}_\mu{}^{\mbox{\rm\scriptsize C}}({\bf x})$ is a part of the gauge
field whose excitation operator content is limited to $a_Q({\bf k})$ and
$a_Q{}^\star({\bf k})$, and its time-dependence in the Heisenberg
picture is trivial. ${\cal A}_i{}^{\mbox{\rm\scriptsize C}}({\bf x})=0$,
${\cal A}_0{}^{\mbox{\rm\scriptsize C}}({\bf x})$ is given by
\begin{equation} {\cal A}_0{}^{\mbox{\rm\scriptsize C}}({\bf
x})=\sum_{\bf k}\frac{1}{\sqrt{k}}\left[a_Q({\bf k})e^{i{\bf k\cdot
x}}+a_Q{}^\star({\bf k})e^{-i{\bf k\cdot x}}\right] \end{equation} In
the Hilbert space appropriate for the common form representation, matrix
elements of ${\cal A}_0{}^{\mbox{\rm\scriptsize C}}({\bf x})$ always
vanish, so that its presence does not interfere with the gauge condition
or the expression for Gauss's law in the Coulomb gauge. When the spinor
field is gauge-transformed, we transform $\tilde{\psi}$, given in
Eq.~(\ref{eq:psitilde}) and represented as \begin{eqnarray}
\tilde{\psi}({\bf x}) &=& \exp\left\{e\sum_{\bf
k}\frac{1}{2k^{3/2}}\left[a_R({\bf k})e^{i{\bf k\cdot x}}-
a_R{}^\star({\bf k})e^{-i{\bf k\cdot x}}\right]\right.\nonumber\\
&+&\left.e\sum_{\bf k}\frac{\phi({\bf k})}{2k^{3/2}}\left[a_Q({\bf
k})e^{i{\bf k\cdot x}}-a_Q{}^\star({\bf k})e^{-i{\bf k\cdot
x}}\right]\right\}\psi({\bf x}). \end{eqnarray} Under a gauge
transformation, $\tilde{\psi}$ transforms as \begin{equation}
\tilde{\psi} \rightarrow \tilde{\psi}^\prime =
\tilde{\psi}\exp\left[ie\chi^{\mbox{\rm\scriptsize cov $\rightarrow$
C}}\right]. \end{equation} The gauge-transformed spinor wave function,
$\tilde{\psi}^\prime$, is most conveniently expressed in the form
$\tilde{\psi}^\prime = \psi + \Upsilon$, where \begin{equation} \Upsilon
= \left\{\exp\left[e\sum_{\bf k}\frac{\phi({\bf k})-
1+\gamma/2}{2k^{3/2}}\,\left(a_Q({\bf k})e^{i{\bf k\cdot x}}-
a_Q{}^\star({\bf k})e^{-i{\bf k\cdot x}}\right)\right]-
1\right\}\psi({\bf x}) \end{equation} and has vanishing matrix elements
in the subspace $\{|n\rangle\}$. $\psi$ is the expression for the spinor
field in the Coulomb gauge.

We can gauge-transform from the covariant to the spatial axial gauge, by
using  $\chi^{\mbox{\rm\scriptsize cov $\rightarrow$ spat}}$ given by
\begin{eqnarray} \chi^{\mbox{\rm\scriptsize cov $\rightarrow$ spat}}&=&
i\sum_{\bf k}\frac{1}{2k^{3/2}}\left[a_R({\bf k})e^{i{\bf k\cdot x}}-
a_R{}^\star({\bf k})e^{-i{\bf k\cdot x}}\right]\nonumber\\ &+& i(1-
\frac{\gamma}{2})\sum_{\bf k}\frac{1}{2k^{3/2}}\left[a_Q({\bf
k})e^{i{\bf k\cdot x}}-a_Q{}^\star({\bf k})e^{-i{\bf k\cdot
x}}\right]\nonumber\\ &+& i\sum_{\bf k}\frac{\epsilon_3{}^n({\bf
k})}{k_3\sqrt{2k}}\left[a_n({\bf k})e^{i{\bf k\cdot x}}-
a_n{}^\dagger({\bf k})e^{-i{\bf k\cdot x}}\right], \end{eqnarray} with
\begin{eqnarray} \partial_i\chi^{\mbox{\rm\scriptsize cov $\rightarrow$
spat}} &=& -\sum_{\bf k}\frac{k_i}{2k^{3/2}}\left[a_R({\bf k})e^{i{\bf
k\cdot x}}+a_R{}^\star({\bf k})e^{-i{\bf k\cdot x}}\right]\nonumber\\
&-& (1-\frac{\gamma}{2})\sum_{\bf k}\frac{k_i}{2k^{3/2}}\left[a_Q({\bf
k})e^{i{\bf k\cdot x}}+a_Q{}^\star({\bf k})e^{-i{\bf k\cdot
x}}\right]\nonumber\\ &-& \sum_{\bf k}\frac{k_i\epsilon_3{}^n({\bf
k})}{k_3\sqrt{2k}}\left[a_n({\bf k})e^{i{\bf k\cdot
x}}+a_n{}^\dagger({\bf k})e^{-i{\bf k\cdot x}}\right] \end{eqnarray} and
\begin{eqnarray} \partial_0\chi^{\mbox{\rm\scriptsize cov $\rightarrow$
spat}} &=& -(1-\frac{\gamma}{2})\sum_{\bf k}\frac{1}{2k^2}j_0({\bf
k})e^{i{\bf k\cdot x}}-\sum_{\bf k}\frac{\phi({\bf k})}{2k^2}j_0({\bf
k})e^{i{\bf k\cdot x}}\nonumber\\ &+&\sum_{\bf
k}\frac{1}{2\sqrt{k}}\left[a_R({\bf k})e^{i{\bf k\cdot
x}}+a_R{}^\star({\bf k})e^{-i{\bf k\cdot x}}\right]\nonumber\\ &+& (1-
\frac{\gamma}{2})\sum_{\bf k}\frac{1}{2\sqrt{k}}\left[a_Q({\bf
k})e^{i{\bf k\cdot x}}+a_Q{}^\star({\bf k})e^{-i{\bf k\cdot
x}}\right]\nonumber\\ &+& \sum_{\bf k}\frac{\epsilon_3{}^n({\bf
k})}{k_3}\sqrt{\frac{k}{2}}\left[a_n({\bf k})e^{i{\bf k\cdot
x}}+a_n{}^\dagger({\bf k})e^{-i{\bf k\cdot x}}\right]. \end{eqnarray} We
observe that \begin{equation} \tilde{A}_\mu{}^{\mbox{\rm\scriptsize
cov}}({\bf x}) - \partial_\mu\chi^{\mbox{\rm\scriptsize cov
$\rightarrow$ spat}} = \bar{A}_\mu{}^{\mbox{\rm\scriptsize spat}}({\bf
x}) + \bar{\cal A}_\mu{}^{\mbox{\rm\scriptsize spat}}({\bf x}),
\end{equation} where $\bar{A}_\mu{}^{\mbox{\rm\scriptsize spat}}({\bf
x})$ is given in Eqs.~(\ref{eq:Aihat}) and (\ref{eq:A0hat}), and
$\bar{\cal A}_\mu{}^{\mbox{\rm\scriptsize spat}}({\bf x})$ has vanishing
matrix elements in $\{|n\rangle\}$.

Finally, the field $\chi^{\mbox{\rm\scriptsize spat $\rightarrow$ C}}$,
given by \begin{equation} \chi^{\mbox{\rm\scriptsize spat $\rightarrow$
C}} = i\sum_{\bf k} \frac{\epsilon_3{}^n({\bf
k})}{k_3\sqrt{2k}}\left[a_n({\bf k})e^{i{\bf k\cdot x}}-
a_n{}^\dagger({\bf k})e^{-i{\bf k\cdot x}}\right], \end{equation} has a
gradient consisting of \begin{equation}
\partial_i\chi^{\mbox{\rm\scriptsize spat $\rightarrow$ C}} = -\sum_{\bf
k} \frac{k_i\epsilon_3{}^n({\bf k})}{k_3\sqrt{2k}}\left[a_n({\bf
k})e^{i{\bf k\cdot x}}+a_n{}^\dagger({\bf k})e^{-i{\bf k\cdot x}}\right]
\end{equation} and $\partial_0\chi^{\mbox{\rm\scriptsize spat
$\rightarrow$ C}}=i[{H}^{\mbox{\rm\scriptsize
C}},\chi^{\mbox{\rm\scriptsize spat $\rightarrow$ C}}]$ given by
\begin{equation} \partial_0\chi^{\mbox{\rm\scriptsize spat $\rightarrow$
C}} = \sum_{\bf k} \frac{\epsilon_3{}^n({\bf
k})}{k_3}\sqrt{\frac{k}{2}}\left[a_n({\bf k})e^{i{\bf k\cdot
x}}+a_n{}^\dagger({\bf k})e^{-i{\bf k\cdot x}}\right]. \end{equation}
With $\chi^{\mbox{\rm\scriptsize spat $\rightarrow$ C}}$ we obtain
\begin{equation} \bar{A}_\mu{}^{\mbox{\rm\scriptsize spat}}({\bf x}) +
\partial_\mu\chi^{\mbox{\rm\scriptsize spat $\rightarrow$ C}} =
A_\mu{}^{\mbox{\rm\scriptsize C}}({\bf x}). \end{equation}

These gauge transformations allow us to connect QED in different gauges.
Not only the gauge fields, but the apparatus of the entire theory can be
transformed in this fashion. We can, for example, start with QED in the
covariant gauge, and express the Hamiltonian in common form either by
substituting the expressions for $\tilde{A}_\mu{}^{\mbox{\rm\scriptsize
cov}}({\bf x})$ in Eqs.~(\ref{eq:AiT}) and (\ref{eq:AiL}), or by
transforming the Hamiltonian given in Eqs.~(\ref{eq:H0cov}) and
(\ref{eq:HIcov}), so that $H^{\mbox{\rm\scriptsize cov}} \rightarrow
\tilde{H}^{\mbox{\rm\scriptsize cov}} = V^{-1}H^{\mbox{\rm\scriptsize
cov}}V$. We can then transform $\tilde{H}^{\mbox{\rm\scriptsize cov}}$
``backwards'' from the common form, treating it now as the spatial axial
gauge Hamiltonian. We transform $\tilde{H}^{\mbox{\rm\scriptsize cov}}
\rightarrow e^{-\Delta}\tilde{H}^{\mbox{\rm\scriptsize cov}}e^{\Delta}$,
and since $\tilde{H}^{\mbox{\rm\scriptsize cov}} =
H_{\mbox{\rm\scriptsize C}} + H_{\mbox{\rm\scriptsize Q}}$, and
$H_{\mbox{\rm\scriptsize Q}}$ commutes with $\Delta$,
$\tilde{H}^{\mbox{\rm\scriptsize cov}} \rightarrow e^{-
\Delta}H_{\mbox{\rm\scriptsize C}}e^\Delta + H_{\mbox{\rm\scriptsize
Q}}$. $e^{-\Delta}H_{\mbox{\rm\scriptsize C}}e^\Delta$ is the
rotationally asymmetric $H^{\mbox{\rm\scriptsize spat}}$.
$H_{\mbox{\rm\scriptsize Q}}$ may either be amputated and discarded, or
retained without affecting the dynamics of state vectors in the quotient
space. To maintain consistency with the common form versions of QED in
all these gauges, $H^{\mbox{\rm\scriptsize cov}}$ and
$H^{\mbox{\rm\scriptsize spat}}$ must operate on quite different Hilbert
spaces. $H^{\mbox{\rm\scriptsize cov}}$ must operate on the space
$\{|\nu\rangle\}$ defined in Sec.~\ref{sec:implementationofconstraints},
while the states on which $H^{\mbox{\rm\scriptsize spat}}$ operates are
the non-normalizable $|\phi_i\rangle$ defined in Eq.~(\ref{eq:phii}).\\

\begin{center} {ACKNOWLEDGMENTS} \end{center}

This research was supported by the Department of Energy under Grant No.
DE-FG02-92ER40716.00.

\newpage \begin{center} Figure Caption \end{center}

\noindent Figure 1.\ \ The Hilbert space for QED in the covariant gauge.
The sheet in the interior of the diagram represents the quotient space
of states $|N_i\rangle$ consisting of electrons, positrons and
transversely polarized photons. The ellipse surrounding the sheet
represents the Hilbert space $\{|n\rangle\}$, in which the states
$|N_i\rangle$ are augmented with zero norm states in which chains of
$a_Q{}^\star({\bf k})$ operators act on $|N_i\rangle$ states. The
rectangle surrounding the ellipse represents the space $\{|h\rangle\}$,
in which the space $\{|n\rangle\}$ is augmented with further states in
which chains of $a_R{}^\star({\bf k})$ and $a_Q{}^\star({\bf k})$
operators act on $|N_i\rangle$ states. The vertical line rising from the
sheet represents a fiber of $|n_i\rangle$ states consisting of a single
$|N_a\rangle$ and the set of all possible zero norm $|n_i\rangle$ states
in which chains of $a_Q{}^\star({\bf k})$ operators act on
$|N_a\rangle$. 
\begin{references} \bibitem{lehmann} H.~Lehmann, K.~Symanzik and
W.~Zimmermann, Nuovo Cimento {\bf 1}, 205 (1955). \bibitem{khallerd18}
K.~Haller, Phys. Rev. D {\bf 18}, 3045 (1978). \bibitem{bjorken} See,
for example, J.~D.~Bjorken and S.~D.~Drell, {\it Relativistic Quantum
Mechanics\/} (McGraw-Hill, New York, 1964). \bibitem{leibbrandt} For a
review of this work, see G. Leibbrandt, Rev. Mod. Phys. {\bf 59}, 1067
(1987). \bibitem{lautrup} B.~Lautrup, Kgl. Danske Videnskab. Selskab,
Mat.-fys. Medd. {\bf 35}, Vol. 11 (1967), sec.~8.2. \bibitem{feynman}
R.~P.~Feynman, Phys. Rev. {\bf 76}, 769 (1949). \bibitem{landau}
L.~D.~Landau and I.~M.~Khalatnikov, Sov. Phys. JETP {\bf 2}, 69 (1956).
\bibitem{fried} H.~M.~Fried and D.~R.~Yennie, Phys. Rev. {\bf 112}, 1391
(1958). \bibitem{tomczak} S.~P.~Tomczak and K.~Haller, Nuovo Cimento
{\bf 8B}, 1 (1975). \bibitem{example} See for example, K.~Gottfried and
V.~F.~Weisskopf, {\it Concepts in Particle Physics\/} Vol. II (Oxford
University Press, 1986), sec. IIB; M.~Kaku, {\it Quantum Field Theory\/}
(Oxford University Press, 1993), chap.~5. \bibitem{gupta} S.~N.~Gupta,
Proc. Phys. Soc. London {\bf 63}, 681 (1950); K. Bleuler,
Helv.~Phys.~Acta {\bf 23}, 567 (1950). \bibitem{landovitz} K.~Haller and
L.~F.~Landovitz, Phys. Rev. D {\bf 2}, 1998 (1970).
\bibitem{haller1}K.~Haller, Acta Phys. Austr. {\bf 42}, 163 (1975).
\bibitem{sohn}K.~Haller and R.~B.~Sohn, Phys. Rev. A {\bf 20}, 1541
(1979). \bibitem{johnson} K.~Johnson and B. Zumino, Phys. Rev. Letters
{\bf 3}, 351 (1959). \bibitem{bialynicki} I.~Bialynicki-Birula, Phys.
Rev. D {\bf 2}, 2877 (1970). \bibitem{kummer} W.~Kummer, Acta Phys.
Austr. {\bf 14}, 149 (1961). \bibitem{haller} K.~Haller, Phys. Lett. B
{\bf  251}, 575 (1990). \bibitem{arnowittt}R.~L.~Arnowitt and
S.~I.~Fickler, Phys. Rev. {\bf 127}, 184 (1962).
\bibitem{dirac}P.~A.~M.~Dirac, {\it Lectures on Quantum Mechanics\/}
(Yeshiva University Press, New York, 1964).
\bibitem{bergmann}P.~G.~Bergmann and I.~Goldberg, Phys. Rev. {\bf 98},
531 (1955). \end{references}
\end{document}